\documentclass[cernpreprint, atlasdraft=false, texlive=2023, UKenglish, texmf, orcidlogo]{atlasdoc}
 
\usepackage[subfigure]{atlaspackage}
\usepackage{atlasbiblatex}
 
\usepackage{atlasphysics}
 
\usepackage{multirow}
 
\graphicspath{{logos/}{figures/}}
 
\usepackage{ANA-IDTR-2022-04-PAPER-defs}

\AtlasTitle{Software Performance of the ATLAS Track Reconstruction for LHC Run 3}
\PreprintIdNumber{CERN-EP-2023-168}
\AtlasJournal{CSBS}
\AtlasJournalRef{Comput Softw Big Sci 8, 9 (2024)}
\AtlasDOI{10.1007/s41781-023-00111-y}
\AtlasAbstract{
Charged particle reconstruction in the presence of many simultaneous proton-proton ($\emph{p}\emph{p}$) collisions in the LHC is a challenging task for the ATLAS experiment's reconstruction software due to the combinatorial complexity.
This paper describes the major changes made to adapt the software to reconstruct high-activity collisions with an average of 50 or more simultaneous $\emph{p}\emph{p}$ interactions per bunch crossing (pile-up) promptly using the available computing resources.
The performance of the key components of the track reconstruction chain and its dependence on pile-up are evaluated, and the improvement achieved compared to the previous software version is quantified.
For events with an average of $60~\emph{p}\emph{p}$ collisions per bunch crossing, the updated track reconstruction is twice as fast as the previous version, without significant reduction in reconstruction efficiency and while reducing the rate of combinatorial fake tracks by more than a factor two.
}
 
\author{The ATLAS Collaboration}
 
\AtlasRefCode{IDTR-2022-04}

\hypersetup{pdftitle={ATLAS document},pdfauthor={The ATLAS Collaboration}}
 
\begin{document}
\crefname{figure}{Figure}{Figures}
\Crefname{figure}{Figure}{Figures}
 
\maketitle

\section{Introduction}
 
The reconstruction of charged particle trajectories (tracking) in the inner detector (ID) and muon spectrometer (MS) is a central part of the ATLAS~\cite{PERF-2007-01,GENR-2019-02} experiment's event reconstruction. During \runTwo of the LHC (2015-2018), ID tracking was the most computing resource intensive component of the offline reconstruction workflow.
Given the proximity to the interaction point (IP) and high granularity of the ID, a single inelastic $\protonProton$ interaction leads to up to 200 individual measurements being recorded in the silicon detectors.
 
During LHC physics data-taking, each proton bunch crossing results in a number of such \protonProton interactions taking place simultaneously (\puText). The expected number of these interactions, $\mu$, is a function of the beam parameters and frequently reported as an average over an interval of data-taking time during which instantaneous luminosity, detector and trigger configuration and data quality conditions are considered constant. These intervals, also referred to as \textit{luminosity blocks} (LB), in general correspond to $60$ seconds of data-taking. In addition to this average value $\avgPileup$, it is also possible to define a value specific to a particular $\protonProton$ interaction event, by accounting for the populations of the individual colliding proton bunches based on their location within the LHC bunch pattern. This more granular estimate of the \puText is referred to as $\pileup$ in the following.
 
In an average ATLAS bunch crossing event with $\avgPileup$ between 10 and 70, up to \num{15000} silicon hits need to be processed, decoded, and combined into clusters. The clusters then need to be combined into short track seeds that the reconstruction subsequently
attempts to extend through the entire ID to identify the charged particles (tracks) and precisely reconstruct their trajectories. The same has to be done for measurements in the MS to identify muon tracks, and the ID and MS tracks have to be combined to obtain final muon candidates.
 
In addition to the growing computational effort, the quality of the track candidates becomes challenging to maintain under high \puText. Firstly, the high density of clusters leads to incorrect cluster-to-track association, potentially degrading the reconstructed track properties relative to their true values. Secondly, collections of unrelated clusters can be combined into false-positive (``fake") tracks, which happens more frequently as the number of available clusters increases with \puText.
 
\begin{figure}[h]
\centering
\includegraphics[width = 0.6 \textwidth]{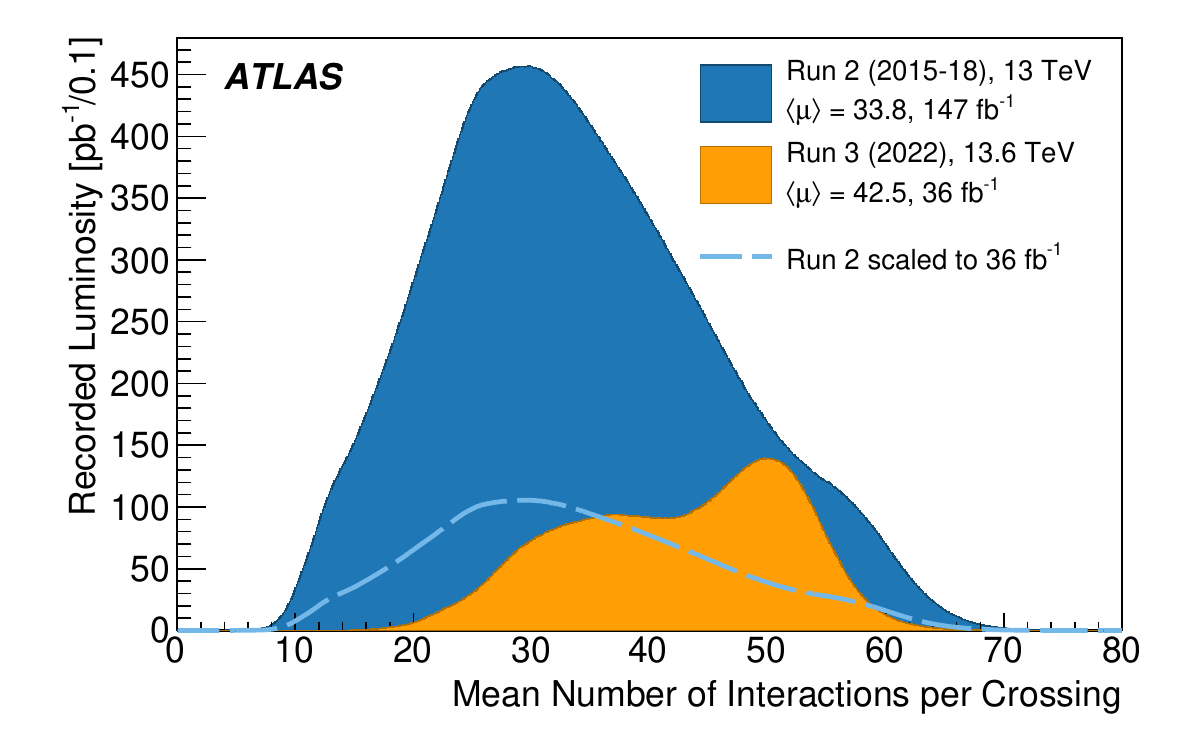}
\caption{
Distribution of the average number of \protonProton collisions at the ATLAS interaction point during the \runTwo and the 2022 \runThree LHC data-taking. The dashed line depicts the \runTwo distribution normalised to the same integral as the \runThree distribution.
\label{fig:muRun23}}
\end{figure}
 
The peak instantaneous luminosity delivered to ATLAS during LHC \runTwo was $1.9\times 10^{34} \text{cm}^{-2}\text{s}^{-1}$~\cite{DAPR-2021-01}, twice the original LHC design value for which the detector and reconstruction software were initially optimized. The ongoing LHC \runThree aims to surpass this value and record up to $300~\text{fb}^{-1}$ of collision data at $\sqrt{s} = 13.6~\TeV$ in the period 2022--2025.
\cref{fig:muRun23} shows the distribution of $\avgPileup$ at the ATLAS interaction point for the \runTwo data-taking compared to 2022, the first year of \runThree . While this value predominantly ranged from 20 to 40 during \runTwo, a large fraction of the \runThree dataset is recorded at 50 or more \protonProton interactions per bunch-crossing, with peak values beyond 60 during the 2022 data-taking, exceeding the average value during LHC \runTwo by a large margin. This is expected to increase further during the following years of \runThree, as the full potential of the LHC is exploited.
 
An extensive set of updates to the ATLAS reconstruction software for \runThree enables the ID track and muon reconstruction algorithms to cope with such conditions.
One key improvement to the entire ATLAS reconstruction for the new run is the adoption of multi-threading to make more efficient use of the available computing resources~\cite{ATL-SOFT-PUB-2021-002}.
In addition to this infrastructural change, a major effort was carried out to improve the per-thread performance of track reconstruction in the ID and MS while maintaining comparable or even superior quality of the reconstructed tracks.
This paper describes the changes made in this effort and demonstrates their impact on the computational and physics performance of track reconstruction in collision data.
 
This paper is structured as follows: the ATLAS detector is described in \cref{sec:atlas2a}, and an overview of the legacy ATLAS track reconstruction chain used in Run 2 is detailed in \cref{sec:reco}. The datasets used and the methology for benchmarking are covered in \cref{sec:methodology}, \cref{sec:octopi} outlines the optimisation procedure, and finally the performance results are delineated in \cref{sec:performance}.


\newcommand{\AtlasCoordFootnote}{
ATLAS uses a right-handed coordinate system with its origin at the nominal interaction point (IP)
in the centre of the detector and the \(z\)-axis along the beam pipe.
The \(x\)-axis points from the IP to the centre of the LHC ring,
and the \(y\)-axis points upwards.
Cylindrical coordinates \((r,\phi)\) are used in the transverse plane,
\(\phi\) being the azimuthal angle around the \(z\)-axis.
The pseudorapidity is defined in terms of the polar angle \(\theta\) as \(\eta = -\ln \tan(\theta/2)\).
Angular distance is measured in units of \(\Delta R \equiv \sqrt{(\Delta\eta)^{2} + (\Delta\phi)^{2}}\).}
 
\section{The ATLAS Detector}
\label{sec:atlas2a}
 
The ATLAS detector~\cite{PERF-2007-01,GENR-2019-02} at the LHC covers nearly the entire solid angle around the collision point.\footnote{\AtlasCoordFootnote}
It consists of an inner tracking detector surrounded by a thin superconducting solenoid, electromagnetic and hadron calorimeters,
and a muon spectrometer incorporating three large superconducting air-core toroidal magnets.
 
\begin{figure}
\centering
\includegraphics[width = 0.9 \textwidth]{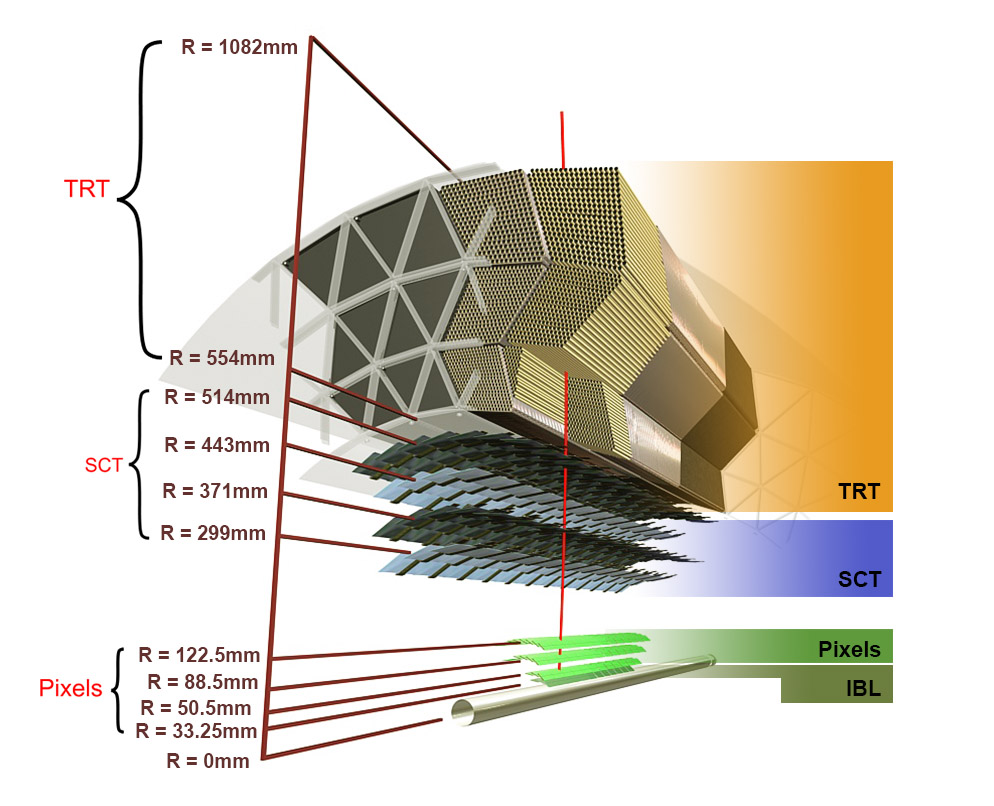}
\caption{Illustration of the barrel region of the ATLAS ID~\cite{PERF-2015-08} as operated during \runTwo and \runThree. \label{fig:IDSketch}}
\end{figure}

The ID is immersed in a \SI{2}{\tesla} axial magnetic field
and provides charged-particle tracking in the range \(|\eta| < 2.5\).
The high-granularity silicon pixel detector is comprised of 4 concentric barrel layers and 3 endcap disks on each side. It covers the vertex region and typically provides four measurements per track, the first hit normally being in the insertable B-layer (IBL), which was installed before Run~2~\cite{ATLAS-TDR-19,PIX-2018-001}.
It is followed by the silicon microstrip tracker (SCT), which is comprised of four barrel layers and 9 planar endcap discs per side, with pairs of modules mounted at a small stereo angle allowing for two measurements per traversed module layer. It usually provides eight measurements per track.
These silicon detectors are complemented by the transition radiation tracker (TRT),
which enables radially extended track reconstruction up to \(|\eta| = 2.0\) and typically records 30 drift time measurements for each track.
The TRT also provides electron identification information
based on the fraction of hits above a higher energy-deposit threshold corresponding to transition radiation. \cref{fig:IDSketch} illustrates the ID detector layout for a quadrant of the full $r$-$z$ plane.
 
The calorimeter system covers the pseudorapidity range \(|\eta| < 4.9\).
Within the region \(|\eta|< 3.2\), electromagnetic calorimetry is provided by barrel and
endcap high-granularity lead/liquid-argon (LAr) calorimeters,
with an additional thin LAr presampler covering \(|\eta| < 1.8\)
to correct for energy loss in material upstream of the calorimeters.
Hadron calorimetry is provided by the steel/scintillator-tile calorimeter,
segmented into three barrel structures within \(|\eta| < 1.7\), and two copper/LAr hadron endcap calorimeters.
The solid angle coverage is completed with forward copper/LAr and tungsten/LAr calorimeter modules
optimised for electromagnetic and hadronic energy measurements respectively.
 
\begin{figure}
\centering
\includegraphics[width = 0.9 \textwidth]{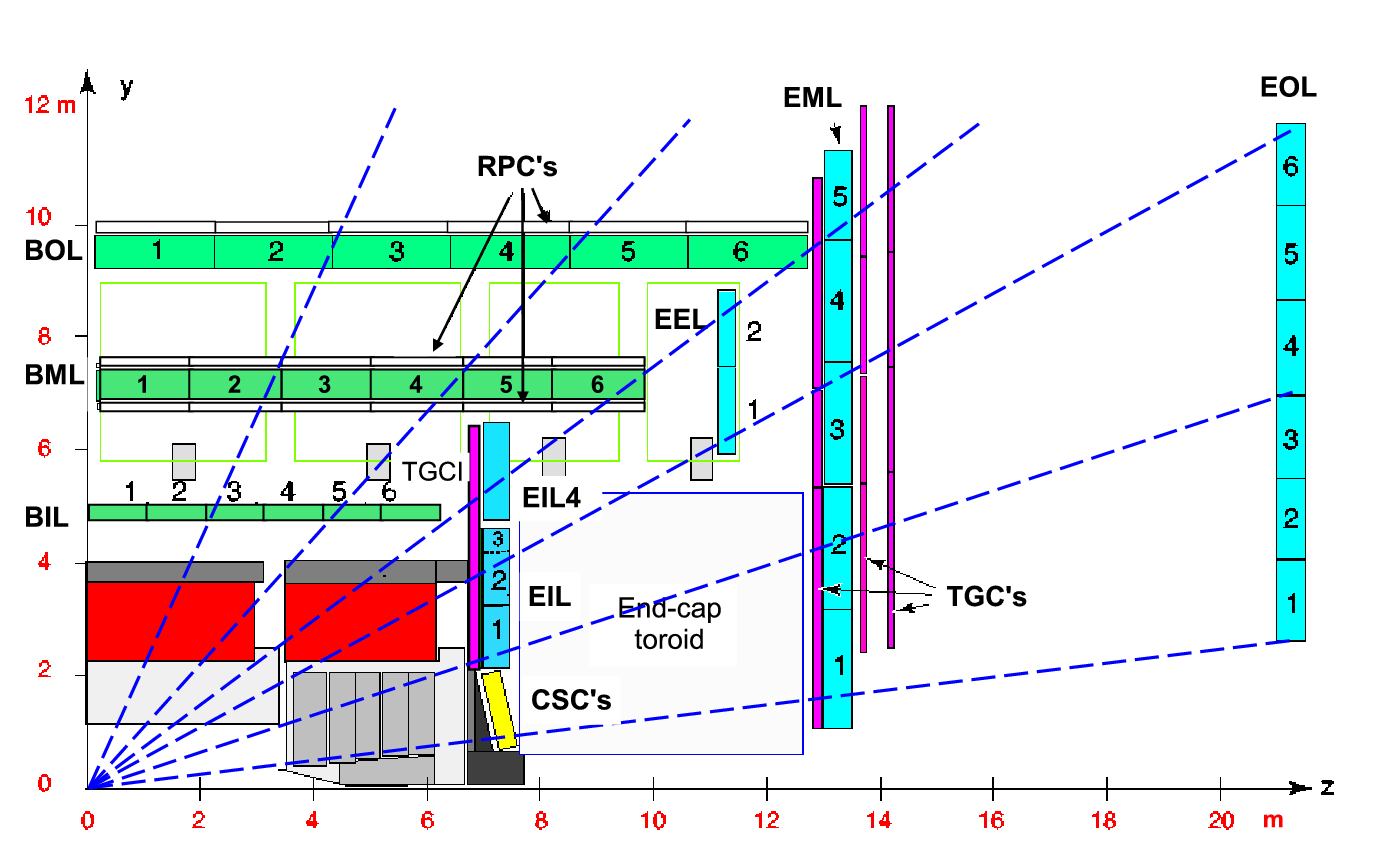}
\caption{ $r$-$z$ view of a quadrant of the ATLAS MS layout~\cite{ATLAS-CONF-2012-125} as operated until \runThree. The green and light blue areas indicate MDT precision tracking chambers, empty and violet boxes indicate RPC and TGC trigger chambers, respectively, and yellow rectangles indicate CSC precision chambers. In \runThree, the innermost endcap layer has been replaced by the New Small Wheel. \label{fig:MSSketch} }
\end{figure}

The MS comprises separate trigger and
high-precision tracking chambers measuring the deflection of muons in a magnetic field generated by the superconducting air-core toroidal magnets.
The field integral of the toroids ranges between \num{2.0} and \SI{6.0}{\tesla\metre}
across most of the detector.
Three layers of precision tracking detectors, each consisting of monitored drift tube (MDT) chambers, cover the region \(|\eta| < 2.7\).
In Runs 1 and 2, cathode-strip chambers (CSCs) took the place of the innermost MDT chambers in the $|\eta| > 2.0$ region, where the background is highest.
The muon trigger system covers the range \(|\eta| < 2.4\) with resistive-plate chambers (RPC) in the barrel and thin-gap chambers (TGC) in the endcap regions.
In the barrel, two RPC chamber layers surround the middle radial MDT layer, forming a common tracking station, and a third RPC chamber layer is located close to the outer MDT layer, forming the outer tracking station. The inner barrel muon station consists exclusively of MDT chambers.
In the endcaps, a layer of TGC chambers is located in front of the inner MDT layer, forming the innermost tracking station together with the CSCs up to \runThree.
One further TGC layer is mounted in front of, and another two behind, the middle endcap MDT layer to form the middle endcap tracking station. The outer endcap  station is comprised of MDT chambers.
In \runThree, to cope with further increasing radiation background, the full innermost layer of the MS endcap detector was replaced by the New Small Wheel, introducing Micromegas (MM) detectors as precision tracking chambers in combination with small-strip Thin Gap Chambers (sTGC)~\cite{ATLAS-TDR-20} while maintaining the same pseudorapidity acceptance of $|\eta| < 2.7$.  The \runTwo MS layout is visualised in the $r$-$z$ plane in~\cref{fig:MSSketch}.
 
Interesting events are selected by the first-level trigger system, which is implemented in custom hardware,
followed by selections made by algorithms implemented in software in the high-level trigger~\cite{TRIG-2016-01}.
The first-level trigger accepts events from the \SI{40}{\MHz} bunch crossings at a rate below \SI{100}{\kHz},
which the high-level trigger further reduces in order to record events to disk at about \SI{1}{\kHz} for \runTwo data-taking.
 
An extensive software suite~\cite{ATL-SOFT-PUB-2021-001} is used in the reconstruction and analysis of real
and simulated data, in detector operations, and in the trigger and data acquisition systems of the experiment.



\section{Legacy Track Reconstruction in ATLAS}
\label{sec:reco}
Particle trajectories are described by five parameters
$\left(d_{0}, z_{0}, \phi, \theta, \frac{q}{p}\right)$
defined with respect to a reference axis,
where $d_{0}$ and $z_{0}$ are the transverse and longitudinal impact parameters, $\phi$ and $\theta$ the azimuthal and polar angle, and $\frac{q}{p}$ the charge divided by the momentum. The default reference for ATLAS tracks is an axis in the global $z$ direction centered on the beam spot. The goal of track reconstruction is to identify sets of electronic signals (``hits``) in the tracking detectors originating from a common charged particle and to estimate the track parameters of the corresponding trajectory.
The collection of hits with its associated track parameter estimate is referred to as a track.
 
The original ATLAS track reconstruction performed during LHC \runTwo is extensively documented and characterised in other sources for both ID~\cite{ATL-PHYS-PUB-2015-051,ATL-PHYS-PUB-2018-002,ATL-PHYS-PUB-2015-031,PERF-2015-08,IDTR-2019-05} and muon~\cite{MUON-2018-03} tracking.
A brief overview of this workflow, also referred to as the \textit{\legText reconstruction}, is provided in the following.
 
\subsection{Inner Detector Tracks}\label{sec:idreco}
 
\begin{figure}
\centering
\includegraphics[width = 0.85\textwidth]{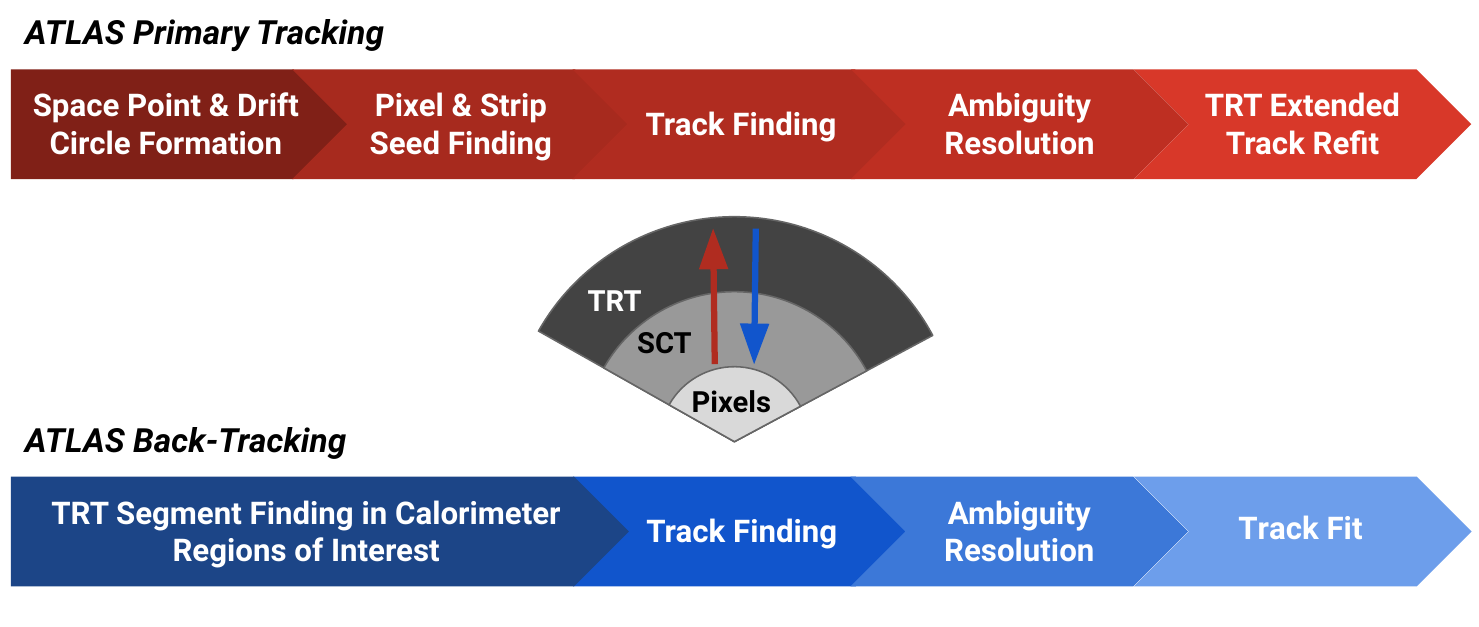}
\caption{Simplified overview of the primary tracking chain and secondary back-tracking chain used in ATLAS ID track reconstruction.
The primary reconstruction runs from inside-out, starting from silicon space-points in the innermost Pixel and SCT subdetectors.
The secondary back-tracking chain runs from outside-in, seeded from leftover TRT hits within electromagnetic calorimeter regions of interest.}
\label{fig:trackingFlowChart}
\end{figure}
The ID track reconstruction is shown schematically in Figure~\ref{fig:trackingFlowChart}. The procedure in the ID starts with a pre-processing stage: signals from adjacent channels in the Pixel and SCT subdetectors are combined into clusters that are interpreted as the deposits left by incident charged particles. Pairs of one-dimensional SCT clusters on either side of a sensor module or individual pixel clusters are then converted into 3-dimensional space-points, with position uncertainties determined by the detector geometry and sensor pitch. In the case of SCT space-points, the missing second coordinate is obtained through the stereo angle between the strips on both sides of the sensor module. ID tracks are reconstructed for $\pT > 500~\MeV$ and $|\eta|<2.5$, where the minimum momentum is driven by the track curvature in the magnetic field and the pseudorapidity interval by the instrumented detector geometry.
 
\subsubsection*{Primary inside-out pass}
 
The primary ATLAS ID track reconstruction starts by forming so-called track seeds, triplets of compatible space-points in the Pixel and SCT subdetectors that could feasibly originate from a single charged particle track produced near the $\protonProton$ interaction point. Search roads (sets of detector modules expected to contain clusters compatible with the seed) in 3D space are built through the remaining detector based on the estimated seed trajectory, and the seeds are extended with additional clusters along the search road into silicon track candidates by means of a combinatorial Kalman Filter \cite{FRUHWIRTH1987444}.
 
To resolve overlaps between track candidates and reject fake tracks, a dedicated ambiguity resolution step is performed.
This scores track candidates based on a range of quality criteria. Lower-quality candidates sharing a large number of associated hits with higher-quality ones are rejected, and the shared hits are assigned to the tracks that are retained.
A limited number of hits shared between different tracks is permitted, to maintain high performance in dense topologies such as cores of high-energy jets, where the separation between charged particles is expected to reach below the magnitude of the sensor pitch.
Clusters determined to consist of more than one charged particle crossing are split among track candidates, with position and uncertainty estimates for each particle crossing provided by a neural network-based algorithm \cite{ATL-PHYS-PUB-2018-002, Khoda:2019xaj} that also assigns a probability for one, two, or more particles to have contributed to the cluster.
 
The refined and purified track candidates determined in the ambiguity resolution step are re-fit using a global $\chi^2$ method to obtain the final, high-precision track parameter estimates. The precision fit runs several iterations until it aborts or achieves the precision required. Aborted fits lead to the track candidates being discarded.
An extension of the track into the TRT subdetector is attempted by collecting drift-circles close to the extrapolated trajectory of the track and performing an iterative re-fit of the entire track including the TRT drift-circles. The additional measurements benefit extended tracks by improving the momentum resolution and allowing for particle identification. To prevent spurious or low-quality extensions which can degrade the momentum resolution, the fraction of so-called \textit{precision hits} is considered. For a hit to be considered a precision hit, the trajectory of the incoming track has to coincide with the measured drift circle radius of the TRT hit within a given threshold expressed as a multiple of the drift radius uncertainty, set to 2.5 in the \legText reconstruction. The precision hit fraction is updated up to the third iteration of the TRT extension fit. For tracks with at least 15 overall TRT hits, at least $30\%$ of the TRT hits along a track have to be classified as precision hits for the TRT extension to be retained. The rate of succesful TRT extensions depends on the subdetector occupancy, but is generally well above $90\%$ for muon tracks that traverse the entire ID.
 
\subsubsection*{Back-tracking and specialised passes}
The primary ID track reconstruction pass is optimised for particles produced at the location of the \protonProton interactions.
To increase acceptance for particles produced at a greater distance from the beamline, such as electrons originating from photon conversions in the detector material, secondary track reconstruction (back-tracking) is performed using the detector hits not already assigned to tracks.
Here, track reconstruction is only attempted in regions of interest determined by deposits in the electromagnetic calorimeter and starts with sets of mutually compatible hits in the TRT ("segments") compatible with the region of interest. In the presence of such a segment, short silicon track seeds consisting of two space-points are constructed in the Pixel and SCT subdetectors. They are extended through the silicon detectors using the same combinatorial Kalman filter used in the primary inside-out pass. A dedicated ambiguity resolution is run upon these track candidates and the resulting tracks are re-fit including TRT extensions.
 
Further dedicated track reconstruction is performed to reconstruct short track segments from muons in $|\eta|>2.5$, where only the pixel detector is traversed, as well as short tracks compatible with decaying, short-lived charged particles. In each case, only leftover hits not used in the prior reconstruction stages are used, to limit combinatorial complexity and avoid track duplication.
 
\subsubsection*{Vertexing}
After the track candidate search has been finalised, the locations of the underlying \protonProton interactions (vertices) are identified by a dedicated vertex reconstruction procedure~\cite{PERF-2015-01}. A first step obtains an initial position estimate for a vertex from the distributions of the $z$ coordinates of closest approach of the tracks to the beamline. Then, a fit of the vertex location is performed, taking into account all tracks loosely compatible with the initial position estimate. Before the changes reported later in this paper, this fit was performed using an iterative procedure, constructing one vertex at a time and removing the associated tracks from consideration before repeating the procedure. The search for secondary vertices from decays of short-lived particles that travel a measurable distance within the tracking volume is not a part of this procedure and instead performed in separate, dedicated downstream reconstruction steps~\cite{ATL-PHYS-PUB-2019-013,ATL-PHYS-PUB-2017-011}, and therefore not included in the optimisation described in this paper.
 
\subsubsection*{Large-$d_{0}$ tracking}
The track reconstruction procedure is optimised to reconstruct tracks that have a small distance of closest approach orthogonal to the beam line (transverse impact parameter, $d_{0}$). Physics analyses making use of  highly displaced tracks, for example to reconstruct decays of long-lived massive particles within the ID volume, need to run an additional track reconstruction pass, referred to as \textit{Large Radius Tracking} (LRT), on hits not used by the previous passes~\cite{LRTPaper}. This pass is similar to the primary inside-out track reconstruction, but configured with a wider search space in the transverse and longitudinal impact parameters, enabling reconstruction of displaced tracks at the price of a drastically slowed execution speed.
In the \legText software,  this additional pass would only be run in a special workflow on a pre-selected sub-set comprising $\mathcal{O}(1\%)$ of the total dataset, to remain computationally feasible. The pre-selection relies on trigger and offline reconstruction without LRT tracks and is driven by the targeted physics analysis signatures.

\subsection{Muon Reconstruction}
 
\begin{figure}
\centering
\includegraphics[width = 0.85\textwidth]{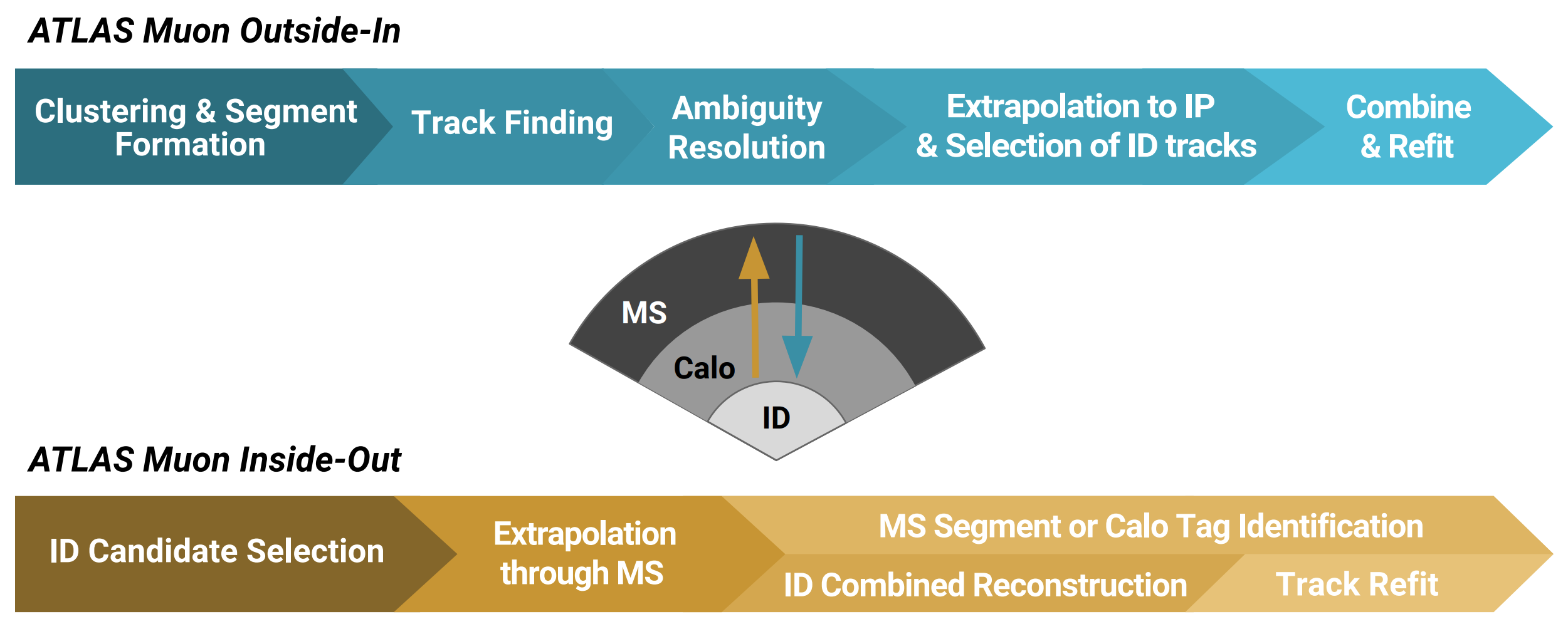}
\caption{Simplified overview of the primary outside--in combined tracking chain and secondary inside--out chain used in ATLAS muon track reconstruction.
The primary chain starts from the MS, building track candidates before attempting to combine them with ID tracks.
The secondary chain starts from ID tracks and attempts to recover muons missed by the primary chain by extending the tracks through the calorimeters and muon spectrometer. They can either be identified as muons using calorimeter- or segment-tagging or extended into inside--out combined muon candidates if sufficient compatible MS measurements are found.\label{fig:muonFlowChart}}
 
\end{figure}
 
Muons are reconstructed using two complementary strategies~\cite{MUON-2018-03}, combining information of the MS with measurements in the inner detector and calorimeters.
The procedure is organised into a pre-processing step followed by two tracking passes - a primary outside--in tracking pass that reconstructs the bulk of the muons, and a secondary inside--out pass that recovers muons in phase-space regions where the outside--in pass is not fully efficient.
\cref{fig:muonFlowChart} illustrates this schematically.
Muons are reconstructed primarily for $\pT > 3~\GeV$~\cite{ATL-PHYS-PUB-2020-002} and $|\eta|<2.7$. The momentum reach is limited by the energy lost while traversing the calorimeters and the pseudorapidity interval by the instrumented muon spectrometer geometry. Muons candidates using ID information are further limited to $|\eta|<2.5$.
\subsubsection*{Pre-processing and segment creation}
In a pre-processing step, signals from adjacent channels in the CSC subdetector are combined into clusters\footnote{This is not needed for the MDT chambers, since muons are not expected to create signals across multiple tubes within the same MDT layer.}. The individual MDT drift circles, CSC, TGC and RPC clusters are collectively referred to as 'hits' or 'measurements' in the following.
ID tracks to be considered for the creation of muon candidates are pre-selected by requiring a minimum quantity of silicon detector hits. Their trajectories are extrapolated outward through the detector to obtain the expected location at which they enter the MS tracking volume, as well as their momentum at this intersection point.
 
The formation of straight line \textit{muon segments} in the MS is initiated in regions of size $0.4 \times 0.4$ in the $\eta$--$\phi$ plane around activity in the trigger chambers. A Hough transform~\cite{Hough2} is used in each MDT chamber and close-by trigger chambers to search for hits aligned on a common trajectory in the bending plane of the detector. The MDT and trigger hits in each muon station are then combined into local muon segments by fitting a straight-line trajectory model to the measurements. The coordinate orthogonal to the bending plane is determined using the trigger chambers if available, while segments comprised of only MDT measurements are assumed to be at the centre of the respective MDT chamber with a large associated uncertainty on the $\phi$ coordinate. A dedicated combinatorial search in the $\eta$ and $\phi$ detector planes is used to identify segments in the CSC chambers in \runOne and \runTwo data.
When building segments, the MDT drift time measurements are calibrated to account for the particle crossing location along the length of the drift tube using the search road position and direction.
 
\subsubsection*{Outside-in reconstruction}
The first reconstruction pass is responsible for reconstructing the bulk of the muons and is referred to as ``outside-in" reconstruction.
It starts by combining the segments across all layers of the MS to form MS-tracks.
The track parameters of a starting \textit{seed} segment are propagated between the MS layers to collect compatible segments, using a parabolic trajectory estimate to account for the bending in the magnetic field at first order.
This procedure is initially seeded from the middle chamber layer with the largest number of available trigger chambers before also including the other layers.
A requirement for the trajectory to point in the approximate direction of the beam line reduces combinatorics, and trigger chamber information is added to obtain three-dimensional trajectory estimates. After this pattern recognition step is complete, a global $\chi^2$ fit of the muon trajectory is performed, taking into account the full track propagation through the inhomogenous magnetic field, possible interactions of the muon with detector material, and the best available knowledge of the real chamber positions obtained using the alignment procedure, exploiting the full $\eta$ and $\phi$ track direction information available at this stage of the reconstruction. Following the initial fit iteration, incompatible measurements  are removed from the track candidate, and compatible measurements along the trajectory that were not assigned to the original candidate are added. The trajectory parameters are refined by a repeated fit to the updated track candidate.
Ambiguities between track candidates are resolved by assigning hits to tracks with higher hit multiplicity and fit quality and rejecting lower-quality tracks that share a large fraction of hits with preferred candidates. To retain efficiency for low-mass dimuon systems, tracks are allowed to share hits in two stations if they do not share hits in a third station.
The MS track trajectories are then extrapolated back to the beam line and re-fit accounting for the muon energy loss inside the calorimeter. The energy loss is estimated using an analytic parameterization derived from a detailed description of the detector geometry combined with the energy measured in the calorimeter. The tracks resulting from this procedure have their track parameters expressed relative to the beam line.
 
MS tracks within $|\eta| < 2.5$ are combined with ID tracks to improve the track parameter resolution.
This procedure starts by collecting ID track candidates around the MS track within an angular compatibility window of size $0.5 \times 1.0$ in the $\eta$ and $\phi$ coordinates expressed at the beam line.
The initial set of candidates is ranked based on a $\chi^2$ comparison between the ID and MS track parameters, and for each candidate a fit of the trajectory to the full set of ID and MS track measurements is attempted, with a global minimization yielding a combined track candidate. The fit accounts for energy loss inside the calorimeter and alignment uncertainties including relative mis-alignment between the ID and MS. MS measurements may be added to or removed from the candidate based on the global fit quality. Finally, the combination with the ID track candidate yielding the best combined fit quality for the given MS track is retained. The \textit{combined muons} obtained in the outside-in chain represent over $95\%$ of the muons reconstructed in ATLAS~\cite{MUON-2018-03}.
In the region $|\eta| > 2.5$, where no full ID tracks are available, MS tracks may be combined with the short pixel track segments discussed in \cref{sec:idreco}, to improve the impact parameter estimate. The resulting subset of the combined muons is called \textit{silicon-associated forward muons}.
If no compatible ID track is found for a given MS track, it is extrapolated to the beam line, taking into account the calorimetric energy loss, and retained as a \textit{standalone muon}. Such muons primarily add acceptance in $|\eta| > 2.5$ for cases where no pixel track segment is available.
 
\subsubsection*{Inside-out recovery}
An inside-out recovery chain complements the outside-in pass to reconstruct muons that have low transverse momenta or enter detector regions where the MS is not fully instrumented.
It comprises three different algorithms: segment-tagged, calorimeter-tagged  and inside-out combined muon reconstruction. All start from ID track candidates and attempt either to identify them as muons or use them as the starting points for combined muon candidates.
To prepare this step, the ID track trajectories are extrapolated through the full MS, recording their predicted crossing location and momentum for any muon chambers they traverse, for use in the inside-out combined reconstruction that will be explained below.
 
\textit{Segment-tagged muon} reconstruction allows ATLAS to recover muons with very low transverse momenta that do not pass beyond the first MS layer. ID tracks are identified as muons through the presence of compatible muon segments.
To find compatible segments, the trajectories of ID tracks are initially extrapolated to a simplified representation of the MS modelled as a series of concentric hollow cylinders representing the barrel layers, and disks representing the endcap layers. For MS segments that cross these layers in locations and with momenta compatible with the ID track, the trajectory of the track is further projected onto the muon chamber of the respective segment, and a more precise compatibility test is performed based on the track and segment directions and their crossing point in the chamber. If at least one compatible segment is found in this way, the track is identified as a muon. In this reconstruction strategy, the MS information is only used for particle identification, and no combined fit is attempted.
 
\textit{Calorimeter-tagged muons} exploit the fact that muons are minimum ionizing particles (MIPs) to recover acceptance in the region $|\eta|<0.1$, where gaps in the MS exist to leave space for calorimeter and ID cabling. Calorimeter clusters along the projected ID track trajectory through the calorimeter are collected and evaluated for compatibility with a MIP signature. If they satisfy this signature, the ID track is identified as a muon, without using information from the MS.
 
The \textit{inside-out combined reconstruction} chain allows to recover combined muons with low transverse momenta that are not found by the outside--in reconstruction. It starts from ID track candidates. The predicted muon chamber crossing locations and directions of their trajectories, as evaluated in the pre-processing phase, are used to build search roads in the MS. Using these roads, a dedicated muon segment-finding pass is performed. The direction and momentum information from the ID-based search roads is used to calibrate the MDT measurements for this phase, allowing to account for the predicted location of the charge deposit along the wire direction when interpreting the drift time measurement. The found segments are collected and added to the track. Using the ID track and the segments recovered in this way, a global re-fit is performed in the same way as for the outside-in chain, to obtain combined muon candidates.


\section{Benchmarking Methodology, Data and Simulation Samples}
\label{sec:methodology}
 
The impact of the software improvements that will be discussed in the following sections on the reconstruction performance is evaluated using collision data and Monte-Carlo (MC) simulated top quark pair production (\ttbar) samples.
 
To evaluate the data reconstruction performance for a given value of \puText in~\cref{sec:performance}, the raw data of a set of 500--1000 consecutive collision events from a common LB are reconstructed, and the processing time taken for the reconstruction as well as the size of the output written to disk are recorded.
This procedure is repeated for sets of events at different \puText values.
The samples are taken from a single LHC fill (fill number 6291) recorded towards the end of the 2017 data-taking campaign and covering a range of \puText values between $\avgPileup=15.5$ and $60$.
In addition, a set of samples from an LHC fill recorded during 2022 as part of \runThree (fill number 8112) with \puText values between $22.4$ and $54.8$ allows ATLAS to compare this information with the \runTwo data ID track reconstruction using the same software.
This ensures consistent data-taking conditions across all \puText values.
Finally, a set of 300 events from LHC fill 5824, recorded in 2017 at $\avgPileup=50$, is used in automated monitoring of ATLAS reconstruction run time on a daily basis. Run times measured in the course of this monitoring are used to evaluate the incremental speed-up of track reconstructions as changes are made in~\cref{sec:octopi}.
All events taken from these two fills fall under the so-called \emph{good-run list} (GRL), meaning that the respective LB satisfy all data quality requirements of the ATLAS physics dataset~\cite{DAPR-2018-01}.
An LHC fill with up to $\avgPileup=90$ recorded in late 2018 (fill 7358) as part of a machine-development campaign is used in addition, to extend the study towards even larger values of \puText.
Unlike the 2017 and 2022 data, this run is not considered part of the ATLAS physics dataset due to the nonstandard data-taking conditions.
Since the ID and MS were fully operational, it is possible to use these events to obtain an estimate of the scaling behaviour of track reconstruction performance under extreme \puText conditions.
\FloatBarrier
 
When evaluating the fraction of successfully reconstructed charged particles, MC simulated \ttbar events are used. The production of these \ttbar events is modelled using the
\POWHEGBOX[v2]~\cite{Frixione:2007nw,Nason:2004rx,Frixione:2007vw,Alioli:2010xd}
generator at next-to-leading order QCD precision with the \NNPDF[3.0nlo]~\cite{Ball:2014uwa} parton distribution function (PDF) set
and the \hdamp parameter\footnote{The
\hdamp parameter is a resummation damping factor and one of the
parameters that controls the matching of \POWHEG matrix elements to
the parton shower, it thus effectively regulates the
high-\pt radiation against which the \ttbar system recoils.} set
to 1.5\,\mtop~\cite{ATL-PHYS-PUB-2016-020}.  The events are interfaced
to \PYTHIA[8.230]~\cite{Sjostrand:2014zea} to model the parton shower,
hadronisation, and underlying event, with parameters set according
to the A14 tune~\cite{ATL-PHYS-PUB-2014-021} and using the \NNPDF[2.3lo]
set of PDFs~\cite{Ball:2012cx}. The decays of bottom and charm hadrons
are performed by \EVTGEN[1.6.0]~\cite{Lange:2001uf}. The effect of \puText was modelled by overlaying the simulated hard-scattering event with inelastic $pp$ events generated with \PYTHIA[8.186]~\cite{Sjostrand:2007gs} using the \NNPDF[2.3lo] set of parton distribution functions~\cite{Ball:2012cx} and the A3 set of tuned parameters~\cite{ATL-PHYS-PUB-2016-017}. For the purpose of studying the reconstruction performance at different levels of \puText, the number of events to overlay for each simulated \ttbar event was sampled from a flat distribution in the interval $\left[0,80\right]$.

All benchmark studies described in this paper were run as the only active user on a dedicated machine
equipped with an AMD~EPYC\texttrademark~7302 16-core processor, running the CERN CENTOS 7 operating system. The processor was operated in ``performance'' mode, with simultaneous multi-threading (SMT) and frequency boosting disabled.
A HS06 score of 22.6 per core was obtained for the processor in this configuration, and a score of 1038 is reported for a dual-processor setup~\cite{HS06} operating with all cores and SMT enabled.
The machine was kept at a stable 50\% in capacity by running an appropriate number of reconstruction tasks simultaneously.
In order to exclude the impact of multi-threading from the comparison, all tests were run in single-thread mode.
 
The executable binary files tested in this work are identical to those most commonly used for the experiment's regular data reconstruction.
As a result, different versions of the software differ not only in terms of
their own programming, but also in terms of the method of compilation as well as
external libraries and the compilation thereof.
All versions of the software were compiled with the default compilation
settings as they were defined in the ATLAS software project at the time of their
release. Since the optimisation flags are partially set according to compiler
presets, the exact details may differ between compiler versions. In all cases,
however, the code is compiled for a generic x86-64 architecture, implying support
for vector instruction set extensions up to SSE2. The binaries produced are therefore
unable to exploit more modern architectural features like AVX.
The binaries for the \legText reconstruction were compiled
using version 6.2.0 of the GNU Compiler Collection, whereas the
updated software release utilises version 11.2.0.
 
The reconstruction run times cited in the following represent the particular set of test events they were evaluated on. Due to statistical fluctuations in the number of charged particles produced per collision event and the particle flux through the various regions of the detector, the run time is expected to vary for evaluations of different sets of events from the same LB. Based on trials performed by sampling different, equal sized sets of events, this variation is found to be on the order of $15\%$ for ID track reconstruction and up to $40\%$ for muon reconstruction, since muons are produced more rarely. However, when evaluating the run time differences between the different software releases, the variation is less than $10\%$ in both cases, as the changes in absolute run time due to different particle multiplicities largely cancel out in this metric.
For a given set of events, the stability of the measured run time across repeated reconstruction runs on the same machine was evaluated to be better than $5\%$.


\section{Software Optimisation}
\label{sec:octopi}
 
A number of changes to the tracking software were introduced in order to optimise the computational performance and the size of the generated output for the conditions expected during LHC \runThree data-taking.
In addition to general algorithmic improvements, the guiding principle is to abort the track reconstruction as early as possible for candidates that are not expected to result in high-quality tracks. This minimises the number of executions of the downstream algorithms, thus saving time and resources.
The experience obtained by operating the reconstruction during \runOne and \runTwo also allows ATLAS to operate the algorithms with less conservative tolerances than were originally implemented, gaining execution speed without sacrificing reconstruction acceptance.
 
\begin{figure}
\centering
\subfigure[ID Tracking]{\includegraphics[width=.49\textwidth]{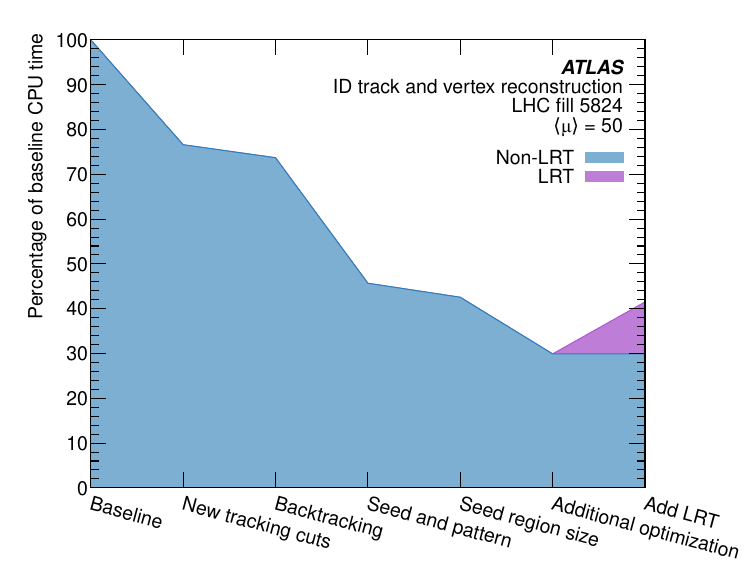}\label{subfig:Progress_INDET}}
\subfigure[Muon Reconstruction]{\includegraphics[width=.49\textwidth]{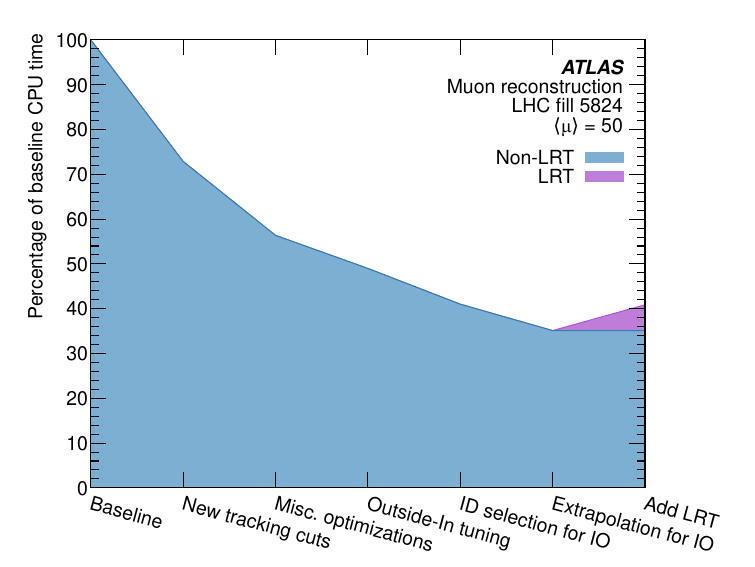}\label{subfig:Progress_MUON}}
\caption{Incremental improvement in execution speed for ID track (a) or muon (b) reconstruction in a set of $\avgPileup=50$ collision data events as improvements are added to the reconstruction software for \runThree. The blue shaded area indicates the time, relative to the initial \runThree software implementation, taken for the track reconstruction. The purple area indicates the time added by the additional LRT. A set of improvements to implementation details of the muon algorithms not discussed in more detail are summarised as ``Misc.\ optimisations'' in (b). Due to changes required to ensure thread-safety, the unoptimised initial performance of the \runThree software, referred to as ``Baseline'', was approximately $20\%$ slower than the previous software used during \runTwo. }
\label{fig:CPU_vs_time}
\end{figure}
 
The reduction in the single-thread CPU timing of ID track and muon reconstruction as  optimisations were incrementally added over time is shown in \cref{fig:CPU_vs_time} for a set of events recorded at $\avgPileup=50$.
The starting point for the optimisations, referred to as "Baseline" in the Figure, is a version of the \legText reconstruction with an initial set of modifications required to ensure thread-safety, and as a consequence slower than the original \legText software implementation described in \cref{sec:reco}. The individual improvements will be detailed in the following.

\subsection{Inner Detector Tracks }
 
The seeding formation time of the inside-out track reconstruction was optimised to prevent seeds unlikely to result in tracks from being passed into downstream processing.
The optimisations include narrowing the acceptance threshold for the estimated impact parameters of SCT seeds from 2~cm to 5~mm, reducing the tolerance in the search road used to extend the seeds from 20 to 12~mm, and preventing Pixel seeds sharing space-points with another seed of lower estimated transverse impact parameter from passing into further processing. The reduced redundancy resulting from the last of these changes was mitigated by exploiting the availability of four pixel layers since the installation of the IBL~\cite{ATLAS-TDR-2010-19,PIX-2018-001} in 2015 by using confirmation space-points to detect promising seeds. A confirmation space-point is a fourth space-point from a different detector layer which, if used to replace the outermost space-point on a given seed, results in a new seed with a curvature compatible with the original, indicating it is likely part of the same charged-particle trajectory. In the updated seed formation procedure, all confirmed seeds are accepted for further processing, even if they share space-points with other seeds. Finally, the window in the transverse impact parameter for accepting seeds dynamically adjusts depending on the deviation of the seed from a straight line in the $r$--$z$ plane, where no deflection due to the magnetic field is expected at the length scale of a seed. The criterion becomes stricter for seeds with a strong deviation from a straight line as these are more likely to be combinatorial artifacts. The introduction of these improvements corresponds to the entry "Seed and pattern" in \cref{subfig:Progress_INDET}.
 
The seeding strategy was further optimised to speed up the seed formation itself, as it contributes a significant fraction of the overall track reconstruction run time due to the combinatorial complexity of the process. This was achieved by adapting the size of the angular regions within which seeds are formed. The regions were reduced to correspond to the track curvature in the magnetic field expected to occur at the lowest track transverse momentum to be reconstructed ($\SI{500}{MeV}$ in the primary ATLAS tracking pass), instead of the wider angular regions used previously. In addition, the knowledge of the detector geometry is applied more rigorously, removing space points from consideration as central point of a seed if they are placed on the outermost or innermost layers of a subdetector or in $\eta$ regions where the subsequent layers provide no acceptance. This improves execution speed without significantly changing the number of tracks being reconstructed. This set of changes is denoted as "Seed region size" in \cref{subfig:Progress_INDET}. Together, the improvements upon the seeding and early track-finding phases reduce the overall ID track reconstruction time by half. They have only a minor impact on the number of correctly reconstructed tracks, while strongly suppressing the occurrence of falsely reconstructed tracks, as will be shown in the following Section.
 
The number of incorrect tracks propagating beyond the initial track-finding stage was further reduced by a re-tuning of the selection criteria applied to the candidates.
Instead of seven silicon clusters per track, at least eight are required in the updated tracking, and the permitted transverse impact parameter range of silicon-seeded tracks is restricted to $|d_0|<5\text{ mm}$ instead of $|d_0|<10\text{ mm}$. This reduces the acceptance of track reconstruction in terms of both displacement and production radius by a small fraction, but significantly lowers the rate of low-quality tracks being reconstructed and written to storage. The reduction in inital track candidates achieved by this optimisation also reduces the number of executions of the subsequent ambiguity resolution procedure and TRT extension phases per event, reducing the track reconstruction time per event by approximately 25\% (as shown in~\cref{fig:CPU_vs_time}). The tuning of the track selection criteria is referred to as "New tracking cuts" in \cref{subfig:Progress_INDET}.
 
The TRT extension was sped up significantly by aborting the iterative track fit procedure early for candidates with insufficient precision hits in the TRT. While the required precision hit fraction of $30\%$ has remained unchanged compared to the \legText reconstruction, the compatibility threshold for precision hits was reduced from 2.5 to 1.75, lowering the fraction of spurious TRT extensions and benefiting momentum resolutions through the reduction of poor-quality TRT extensions. In addition, the iterative fit is now aborted after the third iteration if the final precision hit fraction obtained at this stage does not satisfy the requirement, rather than continuing the fit iterations up to convergence before deciding whether to retain the extension as in the \legText reconstruction. These changes do not impact reconstruction efficiency or the rate of incorrectly reconstructed tracks, but speed up the TRT extension step by nearly $30\%$. In \cref{subfig:Progress_INDET}, the improvements to the TRT extension are included under the entry "Additional optimization".
 
A large number of falsely reconstructed tracks was previously generated by the TRT-seeded back-tracking step. This was reduced by limiting the back-tracking to regions of interest seeded by energy deposits in the electromagnetic calorimeter ($E_{\mathrm{T}} > 6~\GeV$). The recovery of late-appearing tracks from electron conversions, which is the main purpose of this reconstruction step, is degraded at only a negligible level since these topologies coincide with significant calorimetric deposits. However, the number of erroneously reconstructed track candidates is reduced and the execution speed of the back-tracking phase is improved by a factor of 20. This benefits the overall track reconstruction speed on the order of $5\%$ at $\avgPileup=50$, as is visible in~\cref{fig:CPU_vs_time}. These improvements are denoted "Backtracking" in \cref{subfig:Progress_INDET}.  Larger gains are achieved at higher \puText, as the detector occupancy grows.

The iterative vertex finding algorithm (described in Section~\ref{sec:reco}) was replaced by an adaptive multi-vertex fitter algorithm~\cite{ATL-PHYS-PUB-2019-015} in which vertex candidates are allowed to compete for tracks in order to reduce the chance of nearby \protonProton interactions being reconstructed as a single merged vertex. The initial vertex locations are estimated with high accuracy using a Gaussian resolution model for the track impact parameter. This updated algorithm is implemented within the \textit{A Common Tracking Software} (ACTS) framework~\cite{ACTS} which will be the backend of ATLAS track reconstruction during LHC Run 4. This represents the first production use of this framework in an LHC experiment. The improved vertex finding is included under the entry "Additional optimization" in \cref{subfig:Progress_INDET}.
 
Further execution speed was gained by carefully optimising the software implementation of each reconstruction step individually. Notable examples include a re-organisation of the search for holes on tracks performed as part of the precision fit; exploitation of the navigation between detector surfaces already being performed by the track fit procedure;  optimisation of the space-point formation by minimising the number of required matrix multiplications; and the re-writing of parts of the Runge-Kutta propagator implementation used to extrapolate trajectories through the inhomogenous magnetic field of the detector to exploit vectorised instructions where possible. These changes are also part of the entry "Additional optimization" in \cref{subfig:Progress_INDET}.
 
The improvements to the track seeding and finding steps also directly benefitted the LRT reconstruction. Together with further dedicated improvements to the LRT strategy~\cite{LRTPaper}, this made it feasible to integrate fully the LRT pass in the updated ATLAS track reconstruction that is performed for all events.
Removing the need for a custom workflow involving a pre-selection step represents a major extension of the ATLAS physics potential.
In the following study of reconstruction performance, the impact of the LRT step will be pointed out separately, as it was not run by default in the \legText reconstruction.
The impact of adding LRT is shown by a dedicated entry denoted "Add LRT" in \cref{subfig:Progress_INDET}.
 
\subsection{Muon Reconstruction }
 
The muon reconstruction run time was found to scale strongly with the number of ID track candidates considered during the process. The computational effort of both the outside-in and the inside-out passes scales linearly with the number of ID tracks, as the time-consuming combined fits and extrapolations are repeated for each candidate track. The superlinear rate of fake ID tracks with growing \puText in the \legText software was thus directly reflected in a decreasing speed of muon reconstruction.
For this reason, the stricter selection criteria in the improved ID track reconstruction discussed above also benefit the muon reconstruction speed at the order of $30\%$ at high \puText. The impact is visible under the label "New tracking cuts" in \cref{subfig:Progress_MUON}.
 
Before making changes to the muon reconstruction strategy, a series of optimisations to implementation details of the existing algorithms were performed, focusing on memory management and removal of redundant repetition of calculations. These sped up muon reconstruction by around $20\%$ and are referred to as "Misc. optimizations" in \cref{subfig:Progress_MUON}.
 
Dedicated tuning was performed to reduce the number of ID track candidates entering the primary outside-in combined muon reconstruction chain. Following the strategy of aborting reconstruction of non-promising candidates as early as possible, the angular window in the $\eta$--$\phi$ plane applied to select ID tracks for a combined fit was reduced in size from $0.5 \times 1.0$ to $0.2\times0.2$. The use of these stricter tolerances was enabled by comparing the track directions expressed at the MS entrance, rather than the \protonProton interaction region, which increases precision of the MS track parameters. These parameters are already available for ID tracks as a result of the calorimeter extrapolation performed by default on all ATLAS tracks, allowing them to be used without additional overhead. This yields a factor of two speedup of the combined muon reconstruction at $\avgPileup=50$. The introduction of these improvements is labeled as "Outside-in tuning" \cref{subfig:Progress_MUON}.
 
Several further key improvements were made to the inside-out muon chain.
A fundamental change in strategy motivated by the proven reliability and maturity of the combined reconstruction achieved by the end of \runTwo is to no longer execute the inside-out chain independently of the combined reconstruction. The improved inside-out recovery chain only considers ID tracks not already used to form an outside-in combined muon track. This removes the need to extrapolate every ID track candidate through the full MS in preparation for the inside-out combined muon recovery, as $95\%$ of all muons are already found through the outside-in reconstruction.
The ID track extrapolation was also sped up by incrementally propagating the track parameters between adjacent muon chambers, rather than performing a full extrapolation from the MS entrance for each chamber. Together, these changes yielded a $20\%$ speed-up of muon reconstruction.
To further suppress scaling with increasing \puText, ID tracks are considered in the inside-out combined or segment--tagged reconstruction only if hits or segments not belonging to outside--in combined muons exist in the MS chambers towards which their trajectories are oriented. These changes are together shown under the label "ID selection for IO" in \cref{subfig:Progress_MUON}.
 
The segment-tagged muon reconstruction procedure, which in the \legText reconstruction performed extrapolations of all ID tracks to its internal simplified detector geometry in order to match segments to tracks, now instead re-uses the results of the ID track extrapolation to muon chambers performed for the inside-out combined reconstruction. This removes the majority of track extrapolations required during the segment-tagged reconstruction and speeds up the remaining extrapolations for the precise track-segment matching at the level of chambers, because the remaining extrapolation of the ID track to the reference surface of the segments now occurs over a very short distance.
This results in a factor of 15 reduction in processing time for the segment-tagging algorithm, and with the other improvements contributes to an overall speed-up of the inside-out recovery by a factor of five to six in the improved reconstruction. \cref{subfig:Progress_MUON} includes this change under the label "Extrapolation for IO".

Finally, second iterations of both the outside-in combined muon reconstruction and the inside-out recovery steps were added using the set of ID tracks recovered by the LRT step. The algorithmic workflow is identical to the regular muon reconstruction chain, with the exception that no standalone or silicon-associated forward muons are created. This allows the reconstruction of combined and segment-tagged muon candidates with $|d_{0}|$ up to $300$~mm. As this iteration is run independently of the primary muon reconstruction, MS tracks and muon segments may be re-used between the primary and LRT passes. For example, an MS track forming a combined muon with an LRT track may have been reconstructed as a standalone muon in the primary pass. This overlap is resolved in a follow-up step after reconstruction, by preferring combined muons over standalone muons sharing the same MS track and, in the case of two combined muons sharing the same MS track, retaining the one with the better compatibility between ID and MS track directions in the $\eta$ coordinate. This approach was adopted to ensure that the output of the primary muon reconstruction pass remains unchanged if the LRT step is enabled or disabled, a concern during the commissioning of this new feature in reconstruction. The label "Add LRT" indicates the activation of LRT muon reconstruction in \cref{subfig:Progress_MUON}.
 
An additional major effort resulted in the successful inclusion of the sTGC and MM measurements from the New Small Wheel in the reconstruction of \runThree data. This work is however not part of the algorithmic optimisations discussed in this paper and will be reported separately. In the following, muon reconstruction will only be studied on \runTwo data where a meaningful comparison to the \legText reconstruction is possible.


\section{Performance Results}
\label{sec:performance}
 
In the following, the performance of the updated reconstruction is compared to the original, single-threaded software version that was used at the time of data-taking to reconstruct the \runTwo ATLAS dataset. An emphasis is placed on the evolution of the results as a function of \puText.

\begin{figure}[bth!]
\centering
\subfigure[]{\includegraphics[width=.49\textwidth]{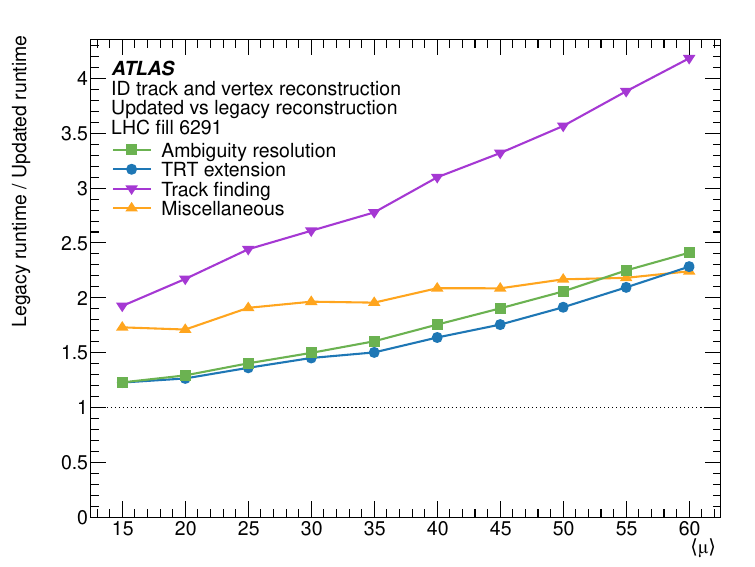}}
\subfigure[]{\includegraphics[width=.49\textwidth]{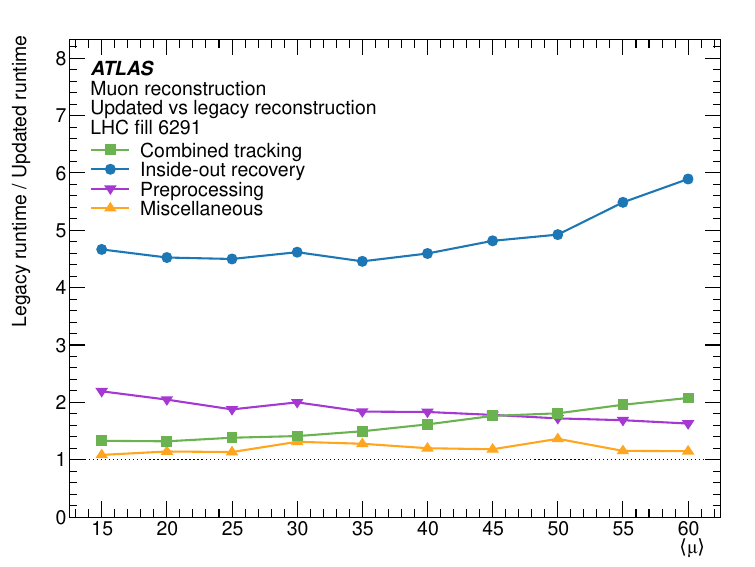}}
\caption{Relative computational performance improvement of key reconstruction steps obtained in the updated reconstruction software compared to that of the \legText implementation when reconstructing a set of LHC \runTwo data events, as a function of $\avgPileup$, for ID track (a) and muon (b) reconstruction. The "Miscellaneous" category includes event data model handling, vertex reconstruction, reading of detector condition information and, for the ID track reconstruction, also cluster formation in the Pixel and SCT subdetectors.  }
\label{fig:CPU_vs_mu_run2_vs_run3_breakdown}
\end{figure}

\cref{fig:CPU_vs_mu_run2_vs_run3_breakdown} shows the relative speedup obtained for the main components of track reconstruction when compared to the original \legText software, as a function of the amount of \puText. The run time of the ID tracking pattern recognition step has been reduced by up to a factor of four, and the other steps of ID tracking are faster by factors of up to two.  This includes drift circle creation, TRT segment making, TRT-seeded back-tracking and TRT extension finding and fitting.
For muon reconstruction, the inside-out recovery step has been significantly sped up. The reduced number of ID track extrapolations performed plays a major part in the factor of five to six run-time reduction for this component. The outside-in combined track muon reconstruction step is also faster by a factor of up to two, profiting from the reduced number of input tracks and the optimised MS--ID track matching, while the pre-processing step has been sped up by a factor of up to two due to a more efficient implementation of the segment formation.
 
\begin{figure}
\centering
\subfigure[]{\includegraphics[width=.49\textwidth,height=0.3\textheight,keepaspectratio]{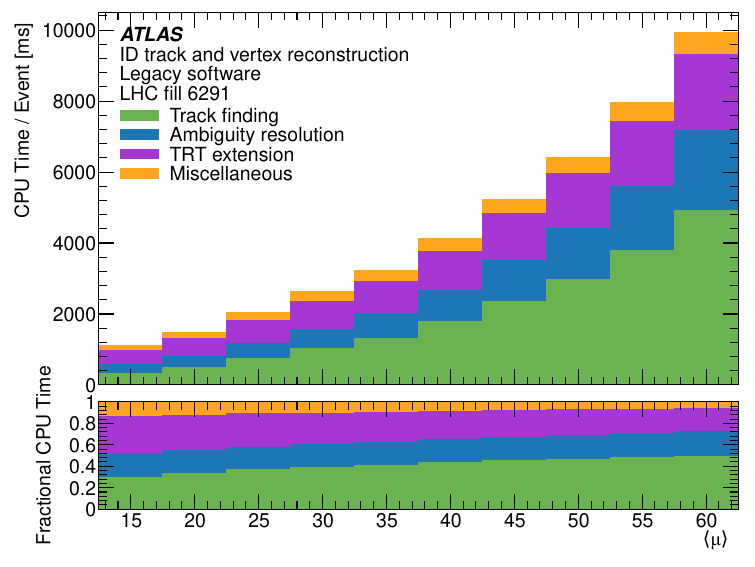}}
\subfigure[]{\includegraphics[width=.49\textwidth,height=0.3\textheight,keepaspectratio]{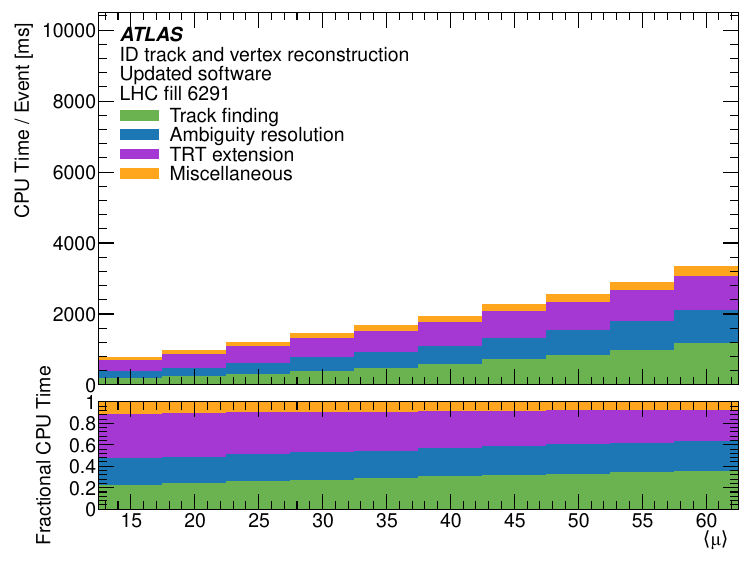}}
\caption{Breakdown of the average CPU time required per event for ID track and vertex reconstruction when reconstructing a set of LHC \runTwo data events, as a function of $\avgPileup$, comparing the \legText (a) and updated (b) configurations. The bottom panel shows the fraction of the total track reconstruction time taken by each component. The "Miscellaneous" category includes event data model handling, vertex reconstruction, reading of detector condition information and, for the ID track reconstruction, also cluster formation in the Pixel and SCT subdetectors.  }
\label{fig:CPU_breakdown_ID_abs}
\end{figure}

\begin{figure}
\centering
\subfigure[]{\includegraphics[width=.49\textwidth,height=0.3\textheight,keepaspectratio]{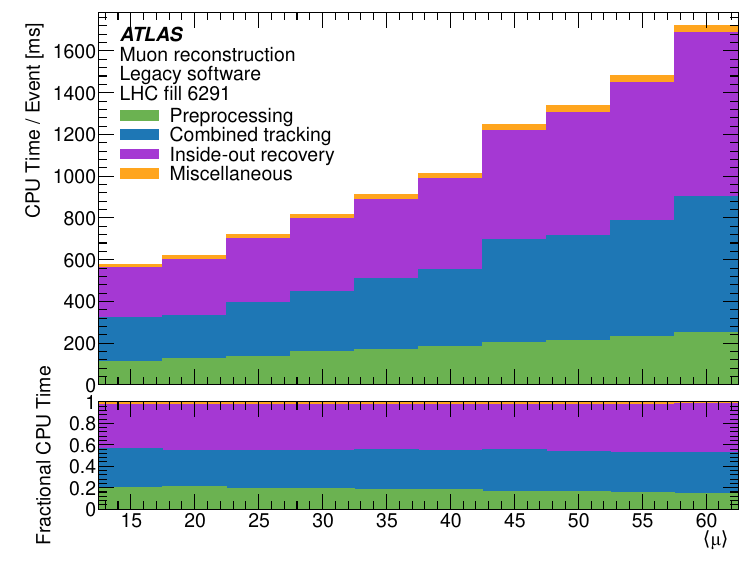}}
\subfigure[]{\includegraphics[width=.49\textwidth,height=0.3\textheight,keepaspectratio]{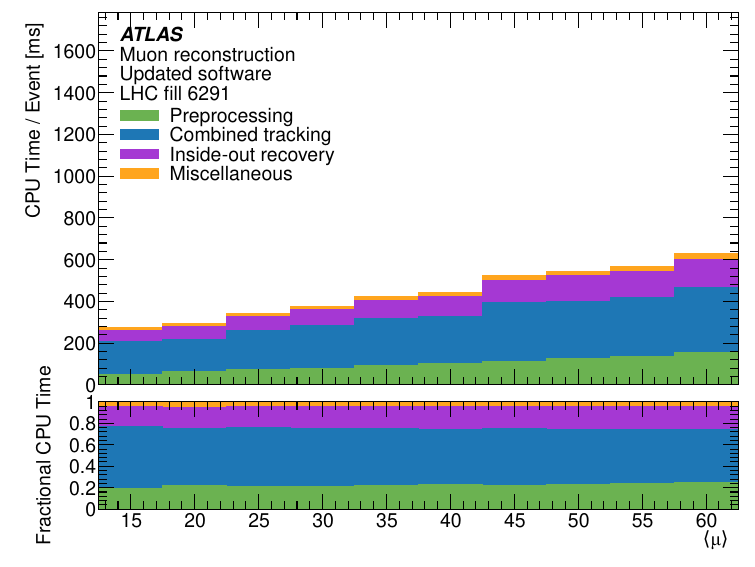}}
\caption{Breakdown of the average CPU time required per event for muon reconstruction when reconstructing a set of LHC \runTwo data events, as a function of $\avgPileup$, for the \legText (a) and updated (b) configurations. The bottom panel shows the fraction of the total track reconstruction time taken by each component. The "Miscellaneous" category includes event data model handling, vertex reconstruction, reading of detector condition information and, for the ID track reconstruction, also cluster formation in the Pixel and SCT subdetectors.  }
\label{fig:CPU_breakdown_MUON_abs}
\end{figure}

A comparison of the average total track reconstruction time per event between the updated and \legText reconstruction software is shown as a function of \puText in \cref{fig:CPU_breakdown_ID_abs} for ID tracks and in \cref{fig:CPU_breakdown_MUON_abs} for muon tracks. The time is decomposed into the key reconstruction steps, and the bottom panels indicate the fraction of the total time contributed by each step. For the \legText reconstruction, the track-finding step was by far the largest CPU consumer for track and total ATLAS reconstruction, and it scaled superlinearly with $\avgPileup$. This behaviour has been rectified in the updated reconstruction. After the optimisations, the pile--up dependency of reconstruction time is closer to linear in the range relevant for \runThree, showing that the updated tracking software is well prepared for high-$\avgPileup$ data-taking. A reduction of the total ID tracking and vertexing CPU time per event by a factor of three is visible for $\avgPileup = 60$, and the CPU time taken is distributed more evenly between the individual components.
For muon reconstruction shown in \cref{fig:CPU_breakdown_MUON_abs}, the strong impact of the improvements to the inside-out muon recovery is immediately visible, driving an overall factor of approximately three speedup of muon reconstruction. In the updated reconstruction, the inside-out recovery, which represented up to $40\%$ of the muon reconstruction time in the \legText reconstruction, contributes only a minor fraction of the total processing time. The majority of the run time is now spent in the outside-in combined reconstruction, which is also responsible for reconstructing $95\%$ of the muon candidates.
 

\begin{figure}
\centering
\subfigure[]{\includegraphics[width=.49\textwidth]{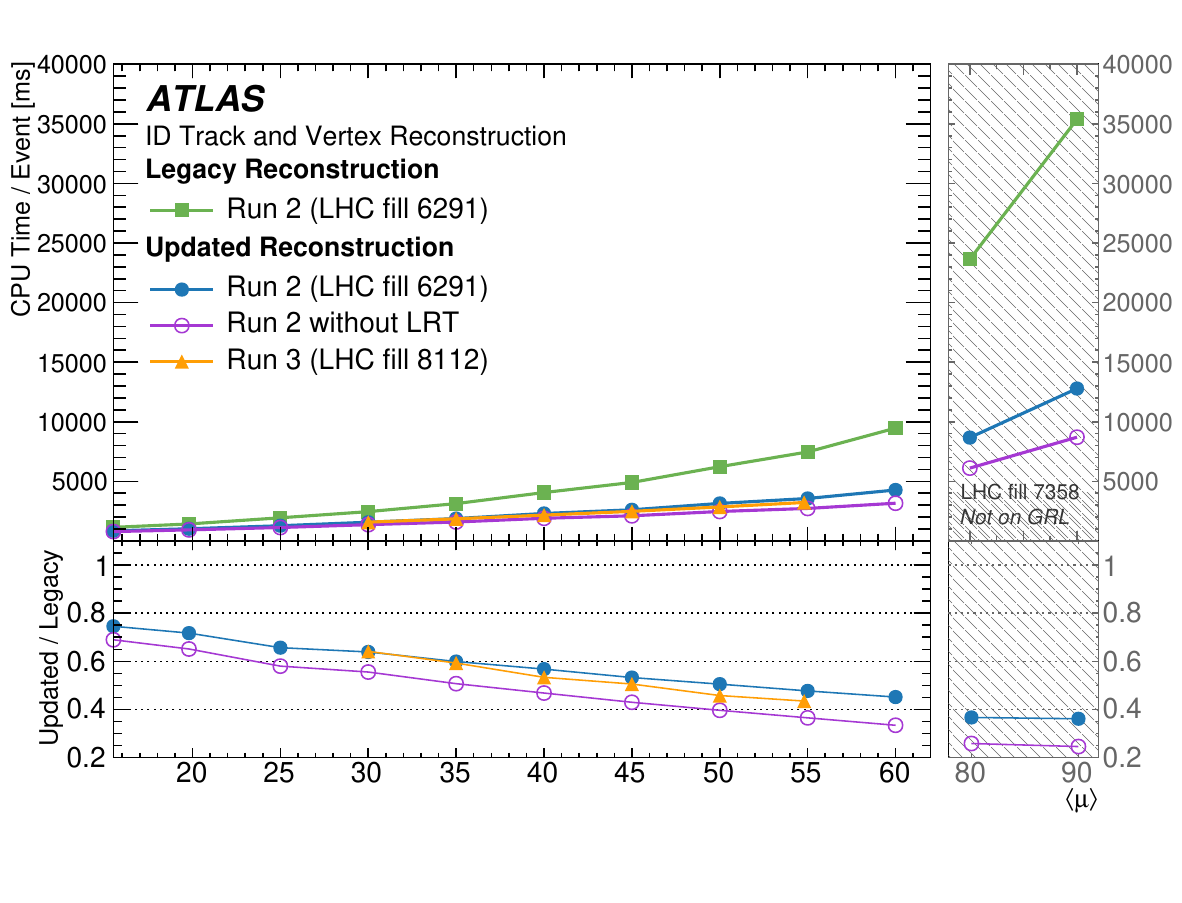}}
\subfigure[]{\includegraphics[width=.49\textwidth]{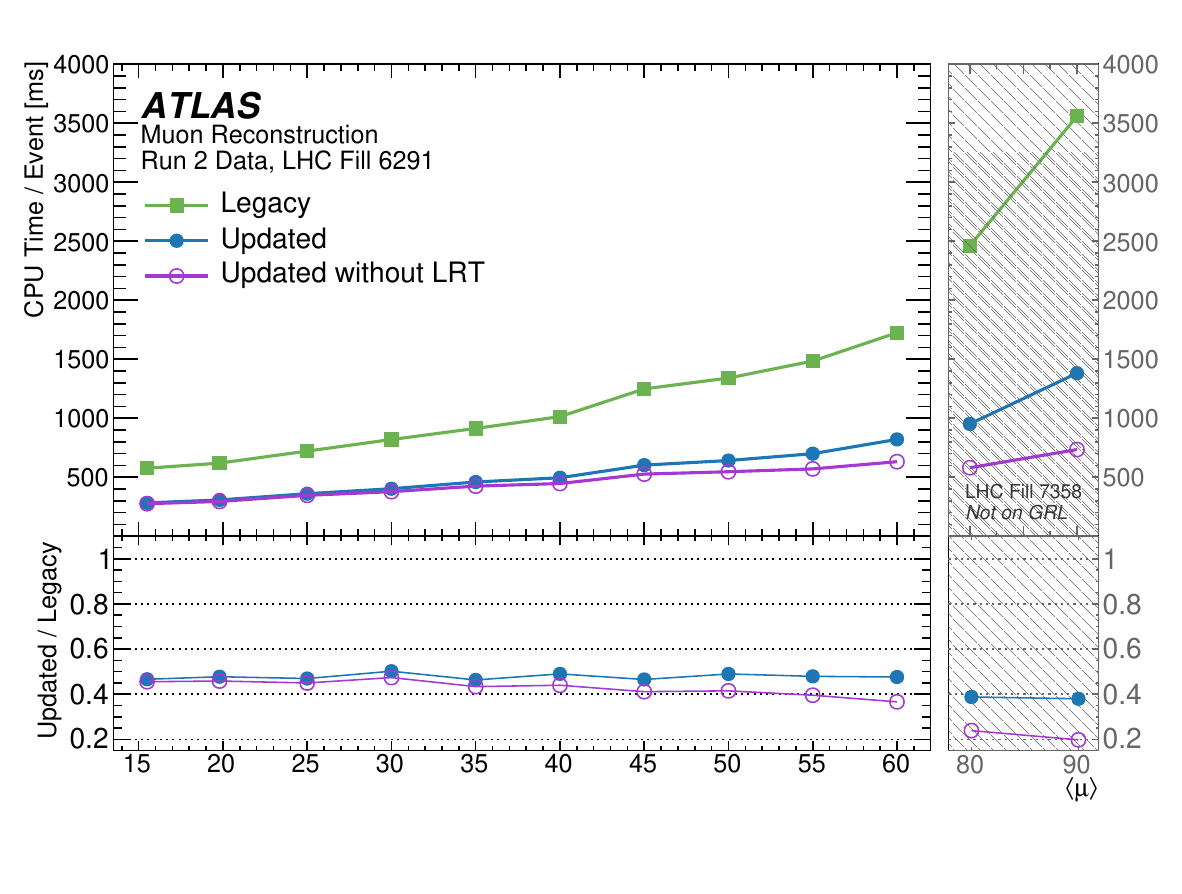}}
 
\caption{Processing time taken per event versus average \puText to reconstruct ID tracks and vertices (a) or muons (b) in the same events, for the the updated (purple) and \legText (green) reconstruction software. The updated reconstruction is also evaluated when including the impact of the additional LRT step (blue). The bottom panel depicts the time taken relative to the \legText reconstruction run time. The shaded area indicates data events taken from a 2018 machine development run not passing the full ATLAS data quality requirements. }
\label{fig:CPU_vs_mu_run2_vs_run3}
\end{figure}

\cref{fig:CPU_vs_mu_run2_vs_run3} shows the total track reconstruction time as a function of $\avgPileup$, including the reconstruction of data taken during an LHC machine-development fill at very high-$\avgPileup$ in 2018.
This allows ATLAS to evaluate the reconstruction speed under such challenging conditions beyond the expected values for \runThree.
For both muon and ID track reconstruction, the updated reconstruction is more than twice as fast as the previous version at $\avgPileup=60$, even after including the new LRT step.
Nearly linear scaling of the CPU consumption with $\avgPileup$ is now observed compared to the behaviour seen for the \legText reconstruction, and even when reconstructing data taken at $\avgPileup=90$, the updated reconstruction achieves a processing speed comparable to the \legText reconstruction at $\avgPileup=60$ when applied to a similar workload.
The benefits from the improvements become more obvious with increasing \puText.
 
For the ID track reconstruction, the CPU time taken to reconstruct a set of \runThree collision data events is also studied, with the LRT step included. The timing of \runThree reconstruction is found to be consistent with the expectation obtained by evaluating the updated reconstruction on \runTwo data. A slightly faster reconstruction compared to the \runTwo reference is due to temporary failures of Pixel detector modules during the 2022 data-taking, which result in a reduced number of space points to process. For muon reconstruction, a direct comparison of reconstruction times between \runTwo and \runThree data is not meaningful as discussed previously.
 
\begin{figure}[tb!]
\centering
\subfigure[]{\includegraphics[width=.5\textwidth]{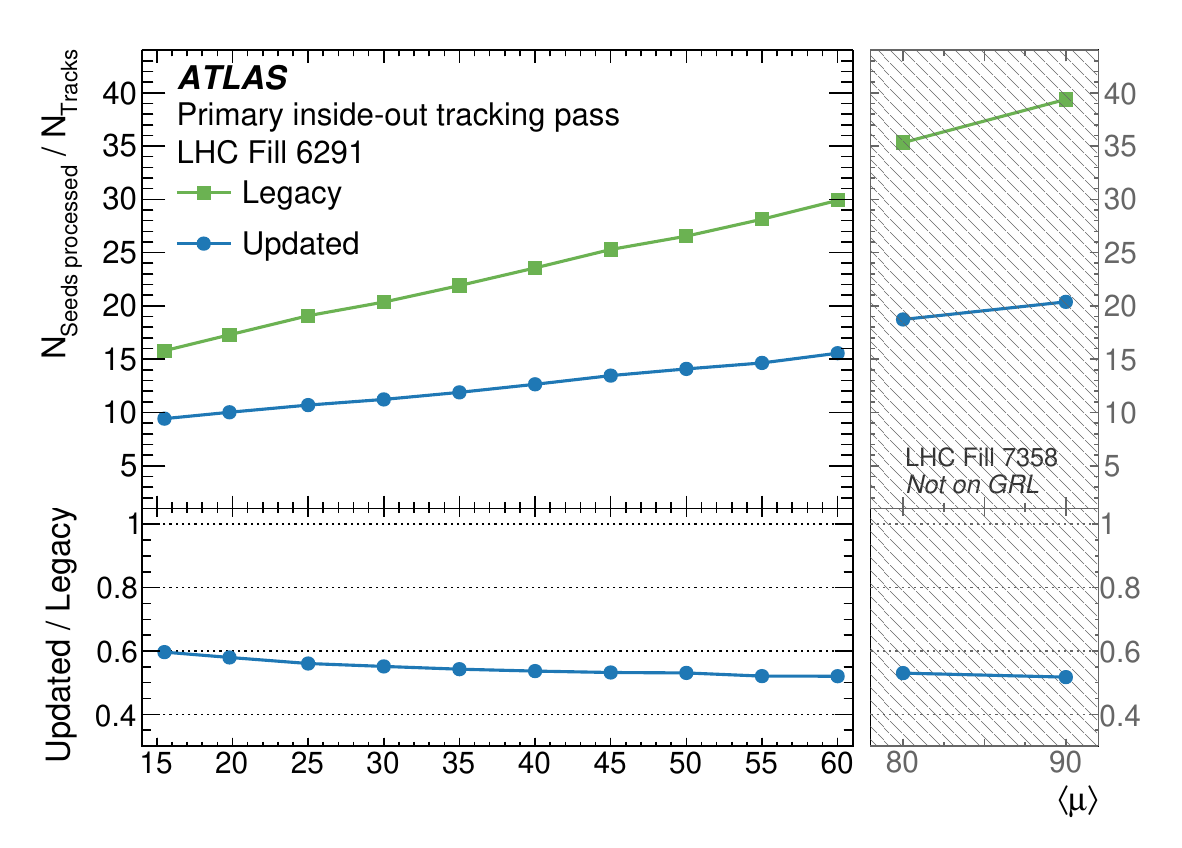}\label{fig:TracksPerSeed}}
\subfigure[]{\includegraphics[width=.48\textwidth]{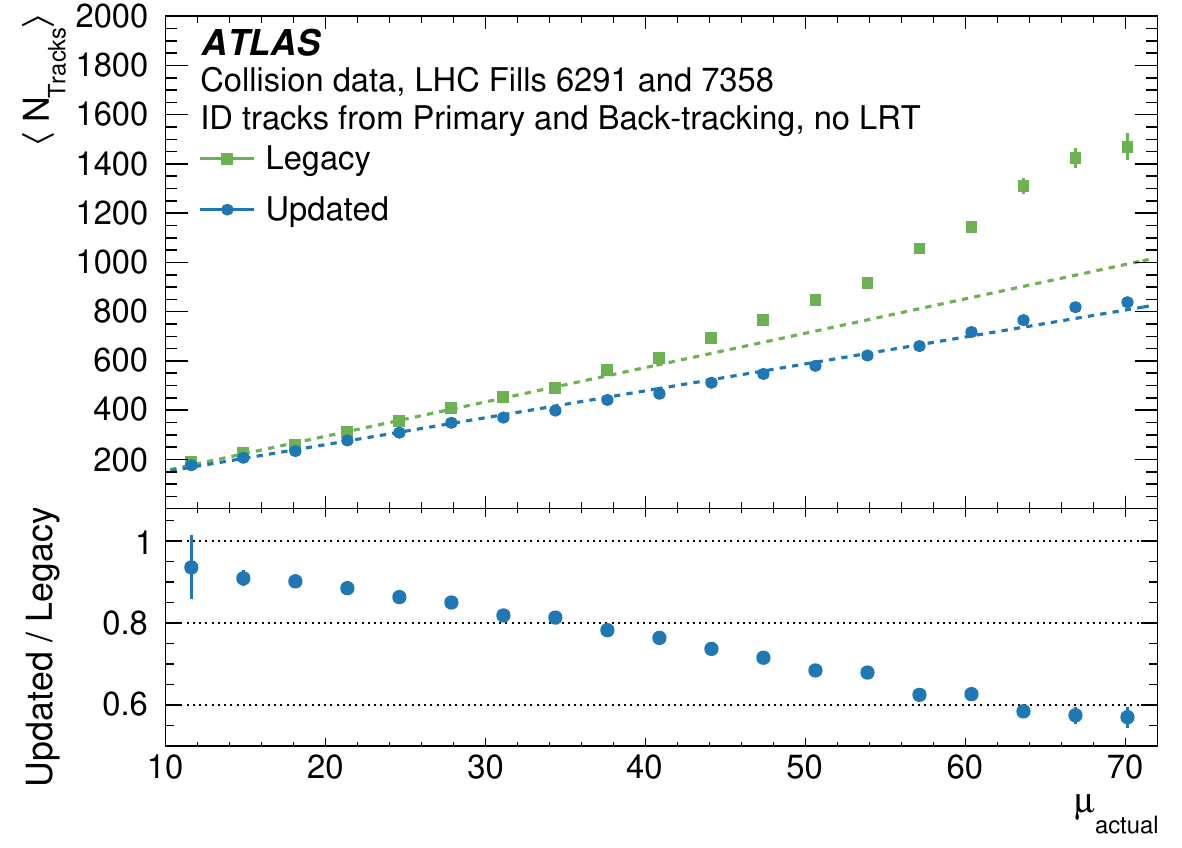}\label{fig:TracksPerEvt}}
 
\caption{Number of seed tracks processed divided by the final number of tracks created in the silicon-seeded tracking as a function of $\avgPileup$ (a) and average number of tracks reconstructed per event as a function of the value $\pileup$, with a linear fit to the range $10 \leq\pileup\leq30$ superimposed as a dashed line (b).}
\label{fig:Number_of_Seeds_Tracks}
\end{figure}

The improved efficiency of the ID track reconstruction is further illustrated in \cref{fig:Number_of_Seeds_Tracks}. \cref{fig:TracksPerSeed} shows the number of processed track seeds divided by the final number of tracks created in the silicon-seeded track finding process as a function of $\avgPileup$. The ratio of seeds to output tracks is reduced by a factor of almost two in the updated reconstruction. To reconstruct the same number of tracks, the costly algorithms responsible for extending a seed through the tracker into a candidate are thus called only half the number of times on average. This allows ATLAS to achieve the large speedup of the track finding stage shown in \cref{fig:CPU_vs_mu_run2_vs_run3_breakdown}.
\cref{fig:TracksPerEvt} depicts the average number of tracks reconstructed per event as a function of $\pileup$.
A linear fit to $10 < \pileup < 30$ is superimposed as a dashed line on the full range of $\pileup$.
The number of genuine track candidates per event is proportional to the number of charged particles produced in the \protonProton collisions and thus scales linearly with $\pileup$.
The number of fake tracks scales superlinearly with $\pileup$, reflecting the growth of combinatorics.
A clear non-linear component amounting to up to $30\%$ of the total number of tracks is visible for the \legText reconstruction, whereas the updated reconstruction shows nearly ideal linear behaviour.
The formation of fake tracks has become very rare in the updated tracking, further improving computational efficiency but also benefiting the reconstruction of track-based physical observables and reducing the disk storage requirements per event.
 
\begin{figure}
\centering
\includegraphics[width=.6\textwidth]{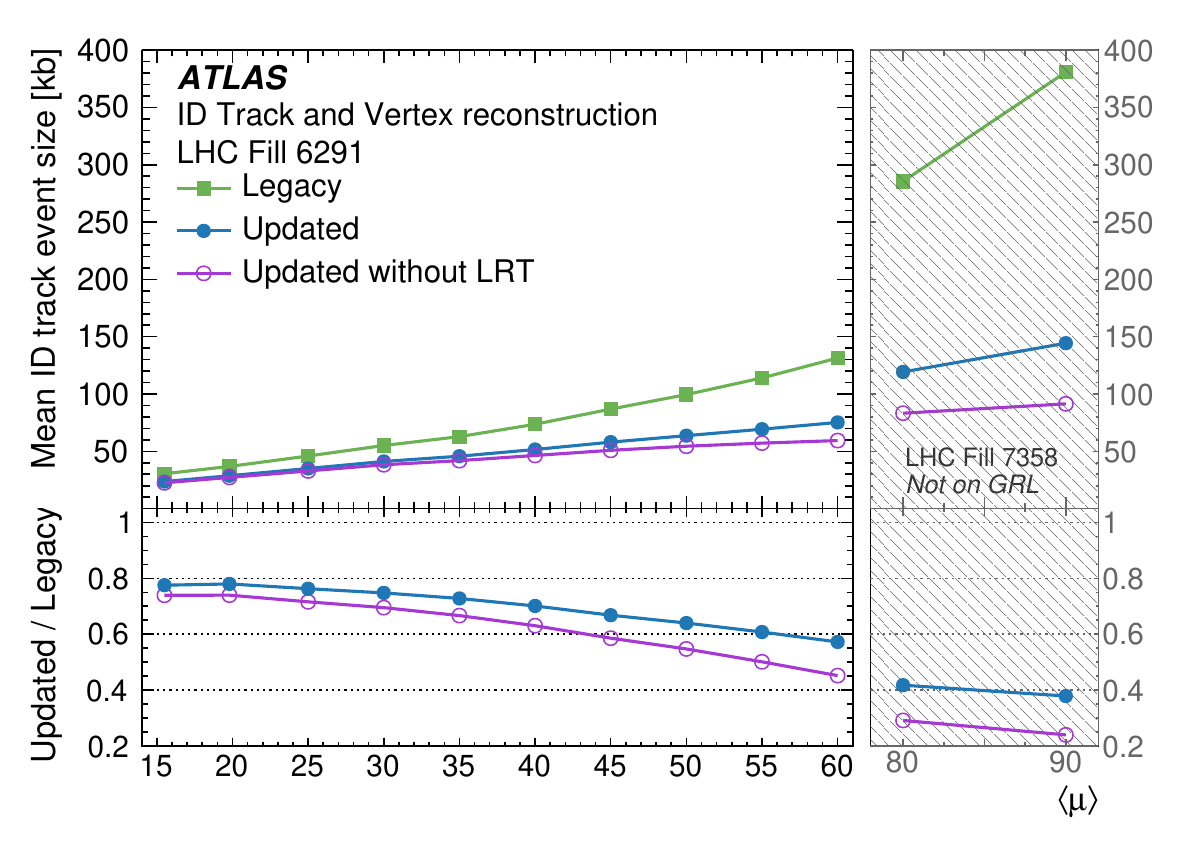}
\caption{Event size of the inner detector reconstruction output in the ATLAS event data format for the same set of reconstructed data events as a function of average \puText, comparing the \legText and updated releases. The shaded panel on the right indicates data events taken from a 2018 machine development run not passing the full ATLAS data quality requirements. }
\label{fig:Size_vs_mu_run2_vs_run3}
\end{figure}
The improvement in data storage requirements resulting from the updated ID track reconstruction is also visible in \cref{fig:Size_vs_mu_run2_vs_run3}, which shows the average disk space required to store ID tracks in kilobytes per event for the standard ATLAS event data format. The observed $20$-$50\%$ reduction compared to the \legText reconstruction is a direct consequence of reducing the rate of misreconstructed ID tracks by a combination of the algorithmic improvements and stricter silicon hit count and impact parameter requirements discussed above.
Even after including the additional tracks from the LRT, a reduction of up to $40\%$ is achieved at $\avgPileup=60$, a value that can be expected for future \runThree data. Additionally, the growth of the disk usage with increasing \puText is observed to be smaller in the updated software release, leading to larger improvements at even higher values of $\avgPileup$.
 
\begin{figure}[tbh!]
\centering
\includegraphics[width=.6\textwidth]{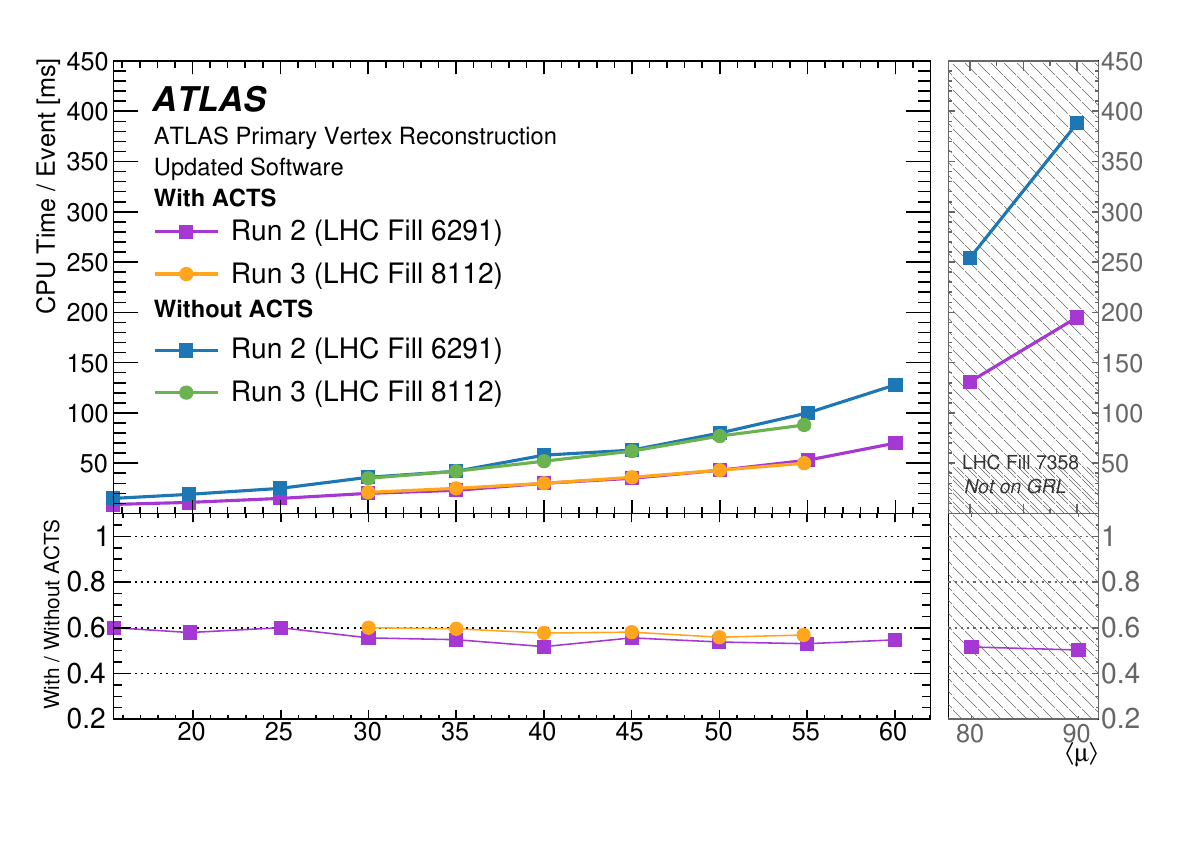}
\caption{A comparison of primary vertex reconstruction processing time taken per event, versus average \puText, for the ATLAS (non-ACTS) version of the Adaptive Multi-Vertex Finder Algorithm and the ACTS-provided implementation. The shaded area indicates data events taken from a 2018 machine development run not passing the full ATLAS data quality requirements.}
\label{fig:CPU_primary_vertexing}
\end{figure}
 
The new deployment of the ACTS software framework in ATLAS for primary vertex reconstruction results in a significant reduction of primary vertexing CPU time. \cref{fig:CPU_primary_vertexing} shows a comparison of the primary vertex reconstruction time required per event between the ACTS-provided Adaptive Multi-Vertex Finder algorithm and the previously deployed ATLAS implementation. An average reduction of CPU time of more than a factor of 1.5 is seen while the ACTS implementation yields physics results that are identical to those from the previous non-ACTS version. The now fully integrated ACTS vertexing software makes additional functionality with further potential for CPU improvements available for future developments.
 
\begin{figure}
\centering
\subfigure[]{\includegraphics[width=.48\textwidth]{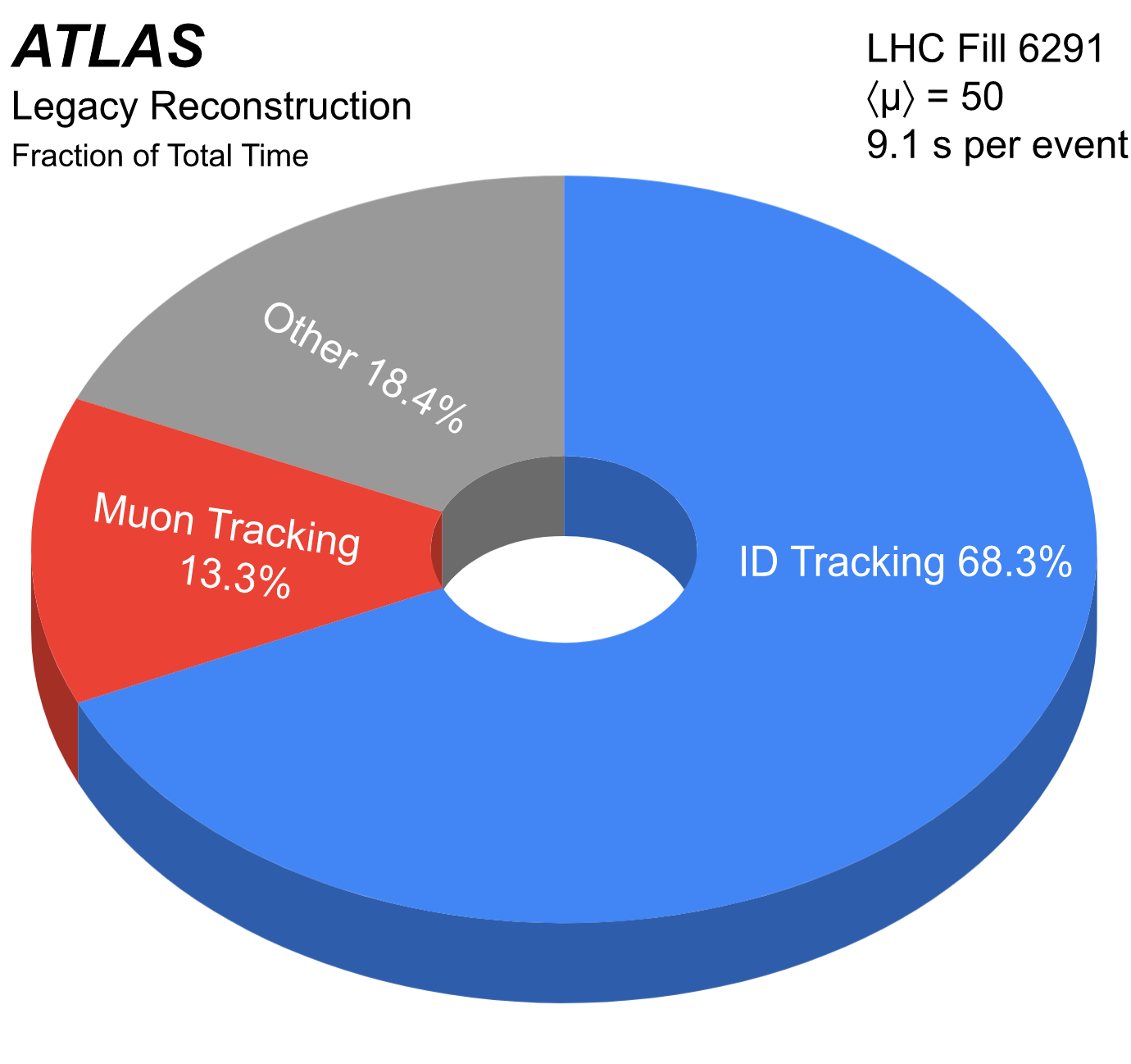}} \hspace{0.02 \textwidth}
\subfigure[]{\includegraphics[width=.48\textwidth]{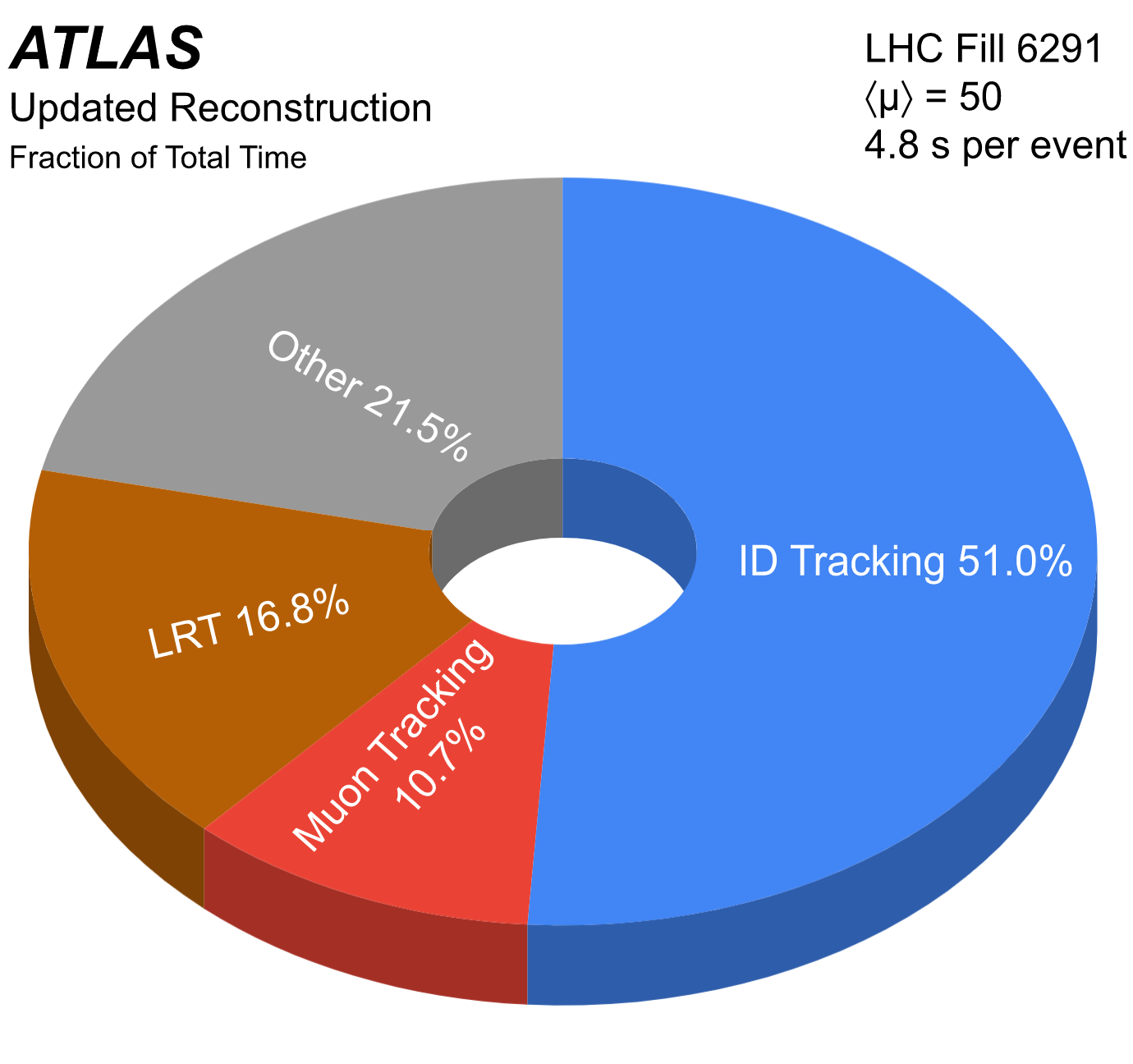}}
\caption{Fraction of the total CPU consumption of full ATLAS reconstruction represented by ID tracking and Muon reconstruction for the \legText (a) and updated (b) reconstruction, for one data run at $\avgPileup = 50$. In the updated reconstruction, several algorithms, such as jet reconstruction and hadronic tau decay candidate identification, are no longer included in the main ATLAS reconstruction and hence no longer enter the sum.}
\label{fig:CPU_pie}
\end{figure}

\cref{fig:CPU_pie} shows that ID track and Muon reconstruction accounted for around $68\%$ and $13\%$ of the total ATLAS event reconstruction CPU time in the \legText software release, respectively. As a result, the speed improvements described in this paper scale almost directly to an improvement of the total ATLAS event reconstruction.
In the updated reconstruction, this fraction is approximately retained when including the LRT.
This is because several other algorithms, such as jet reconstruction and hadronic tau decay candidate identification, which contributed around $3\%$ of the total reconstruction time, have been moved from the main reconstruction pass into the downstream processing in the updated ATLAS event reconstruction. This reduction in runtime of non-tracking algorithms in the main reconstruction step is almost proportional to the speedup achieved in tracking. In addition, a large number of the most time-consuming non-tracking algorithms, such as streaming the reconstructed event data to disk, scale in run time with the number of reconstructed track candidates and thus benefit from the reduced rate of misreconstructed tracks. Accordingly, the total reconstruction time per event has decreased from 9.1s to 4.8s on the machine specified in~\cref{sec:methodology}. This indicates that track reconstruction remains the most promising part of event reconstruction to achieve further improvements in overall throughput.

\begin{figure}
\centering
\includegraphics[width=.6\textwidth]{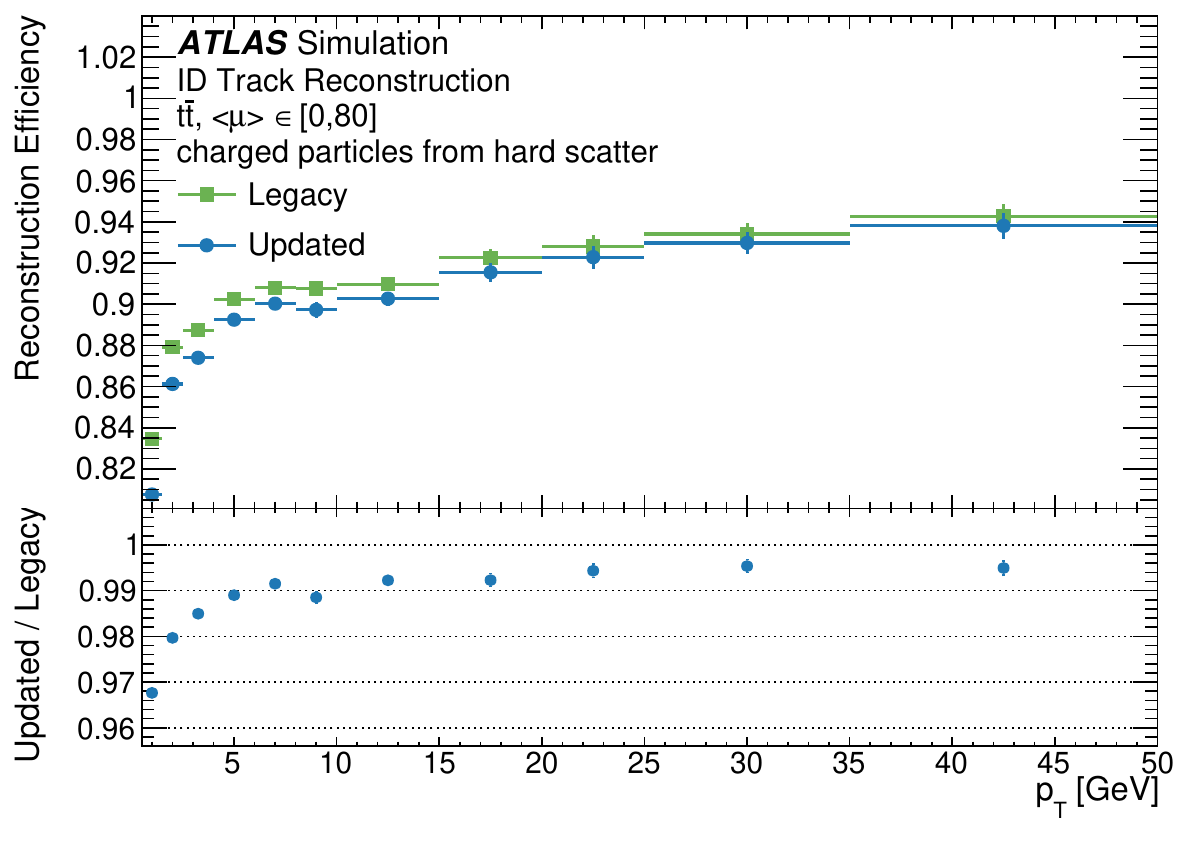}
\caption{ID Tracking efficiency as a function of $\pT$ in simulated $t\bar{t}$ events with a flat $\avgPileup$ distribution between 0 and 80, in a comparison of the updated and \legText reconstructions. The error bars indicate statistical uncertainties. }
\label{fig:Tracking_eff}
\end{figure}

\begin{figure}
\centering
\subfigure[]{\includegraphics[width=.49\textwidth]{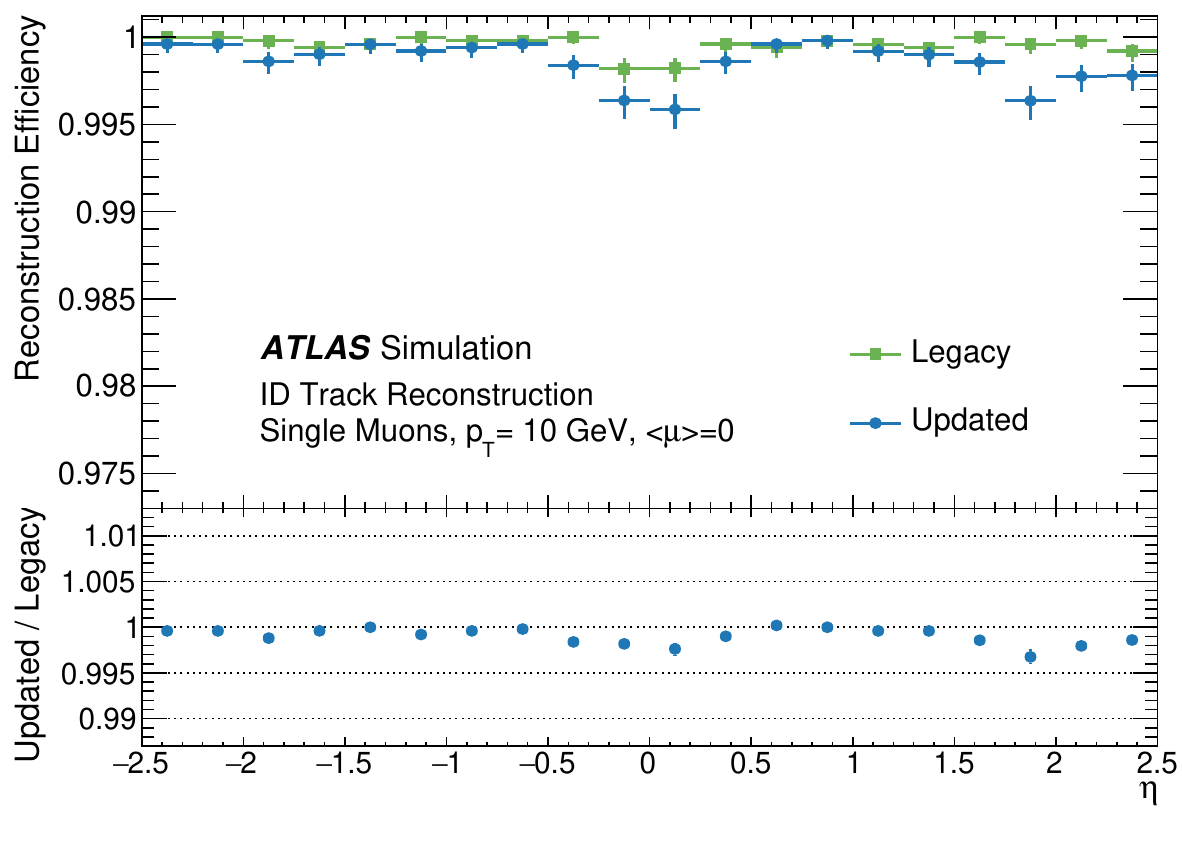}}
\subfigure[]{\includegraphics[width=.49\textwidth]{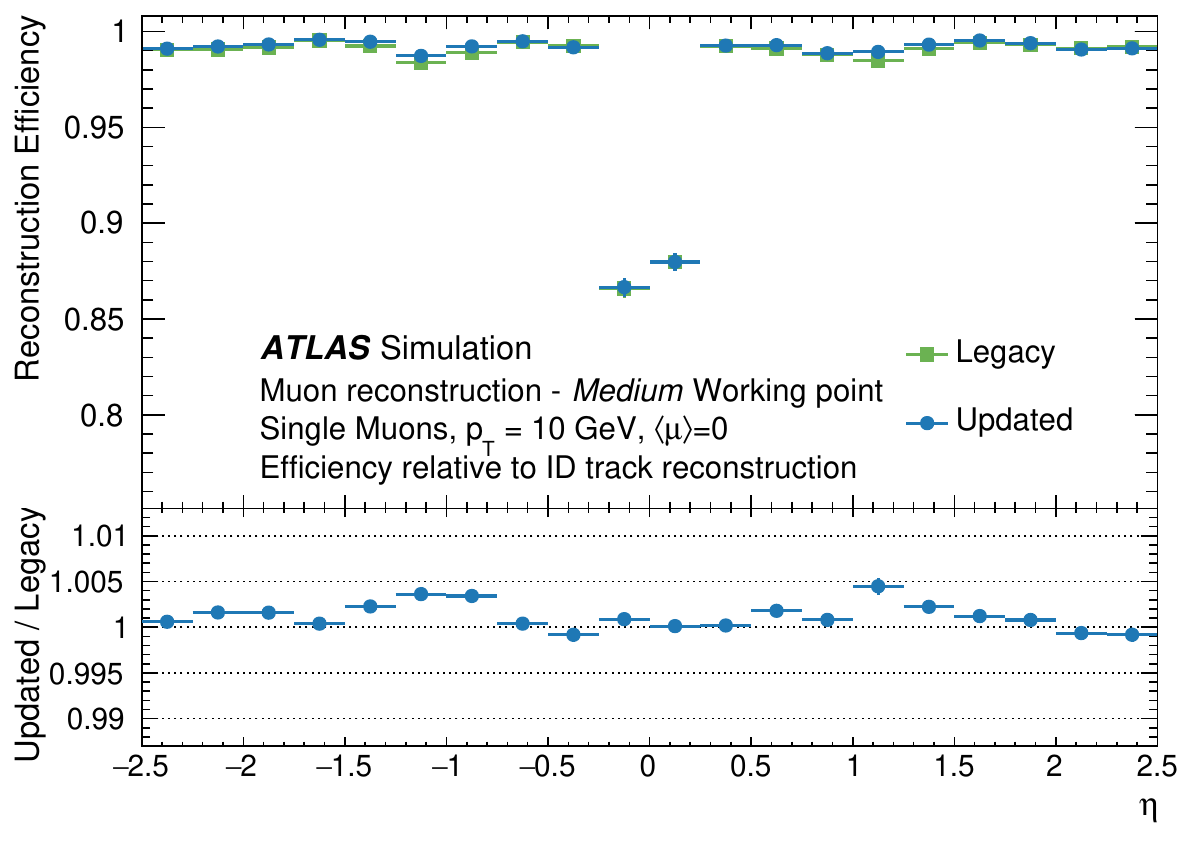}}
\caption{Tracking efficiency for simulated single muons using the \runTwo detector geometry with $\pT = 10~\GeV{}$ for ID track (a) or muon reconstruction using the \textit{Medium} working point in case an ID track was successfully reconstructed (b), as a function of  $\eta$, in a comparison of the updated and \legText reconstructions. The error bars indicate statistical uncertainties. }
\label{fig:MuonEff}
\end{figure}

Finally, the software improvements have not negatively impacted the track reconstruction physics performance compared to the \legText reconstruction.
The tracking efficiency, defined as the fraction of charged particles originating from the primary \protonProton interaction successfully reconstructed, is shown for the ID comparing the updated and \legText reconstruction in \cref{fig:Tracking_eff} as a function of the charged particle transverse momentum. Simulated $t\bar{t}$ events are studied, leading to a track sample dominated by pions. The efficiency loss is smaller than $4\%$ at low $\pT$ and smaller than $1\%$ at larger transverse momenta. The slight reduction compared to the \legText implementation is a result of the stricter requirements on the number of silicon clusters and impact parameter for a track to be retained, and does not affect the reconstruction of muons as minimum ionising particles.
\cref{fig:MuonEff} depicts the reconstruction efficiency obtained for simulated single muons of $\pT = 10~\GeV$ for ID tracks and muons satisfying the \textit{Medium}~\cite{MUON-2018-03} identification criteria, which preferentially select combined muons and represent the requirements most frequently applied in ATLAS physics analyses. The muon reconstruction efficiency is shown relative to the successful identification of an ID track, making the total muon reconstruction efficiency the product of the two values depicted individually.  The updated ID track reconstruction achieves an efficiency within five per-mille of the \legText reconstruction across the full instrumented pseudorapidity range. The muon reconstruction efficiency, once an ID track is found, is slightly improved compared to the \legText implementation. In $|\eta|<0.1$, the muon reconstruction efficiency is reduced for both software releases due to limited detector acceptance, as the MS has gaps for service cabling in this region.
In both figures, the simulations include the best available knowledge of inoperative detector modules. In the improved reconstruction, this includes a slightly larger number of defects as compared to the \legText software implementation, reflecting the updated detector status after the conclusion of LHC \runTwo. This more conservative estimate of detector availability is part of the reason for the small difference between the software releases.
 
\FloatBarrier


\section{Conclusions}
 
LHC \runThree challenges experiments' event reconstruction with a significant increase in the amount of \puText activity.
In ATLAS, the inner detector and muon track reconstruction software has been re-optimized in order to meet this challenge, ensuring that prompt event reconstruction remains feasible within the available computing resources. The improvements exploit the increased experience in operating the experiment by employing stricter tolerances where possible and relying more on the sharing of information between previously independent algorithms. Non-promising track and muon candidates are rejected as early as possible, minimizing the number of calls to downstream algorithms within the reconstruction chain. An execution speed improvement between a factor of 2 and 4, depending on the \puText, is achieved for both ID track and muon reconstruction. The number of fake tracks has been reduced to less than half of the previous amount and up to $40\%$ less storage space is required for the tracks. At the same time, the reconstruction efficiency is nearly completely retained. The performance and reconstruction speed remain stable at \puText values well beyond the design values for LHC \runThree, giving confidence in the robustness of ATLAS event reconstruction to evolving operating conditions. The improvements allow the execution of an additional reconstruction pass to recover tracks not pointing towards the beamline in the standard reconstruction, benefiting long-lived particle searches while retaining a significant overall performance improvement compared to the past software version.


\section*{Acknowledgements}


We thank CERN for the very successful operation of the LHC, as well as the
support staff from our institutions without whom ATLAS could not be
operated efficiently.
 
We acknowledge the support of
ANPCyT, Argentina;
YerPhI, Armenia;
ARC, Australia;
BMWFW and FWF, Austria;
ANAS, Azerbaijan;
CNPq and FAPESP, Brazil;
NSERC, NRC and CFI, Canada;
CERN;
ANID, Chile;
CAS, MOST and NSFC, China;
Minciencias, Colombia;
MEYS CR, Czech Republic;
DNRF and DNSRC, Denmark;
IN2P3-CNRS and CEA-DRF/IRFU, France;
SRNSFG, Georgia;
BMBF, HGF and MPG, Germany;
GSRI, Greece;
RGC and Hong Kong SAR, China;
ISF and Benoziyo Center, Israel;
INFN, Italy;
MEXT and JSPS, Japan;
CNRST, Morocco;
NWO, Netherlands;
RCN, Norway;
MEiN, Poland;
FCT, Portugal;
MNE/IFA, Romania;
MESTD, Serbia;
MSSR, Slovakia;
ARRS and MIZ\v{S}, Slovenia;
DSI/NRF, South Africa;
MICINN, Spain;
SRC and Wallenberg Foundation, Sweden;
SERI, SNSF and Cantons of Bern and Geneva, Switzerland;
MOST, Taiwan;
TENMAK, T\"urkiye;
STFC, United Kingdom;
DOE and NSF, United States of America.
In addition, individual groups and members have received support from
BCKDF, CANARIE, Compute Canada and CRC, Canada;
PRIMUS 21/SCI/017 and UNCE SCI/013, Czech Republic;
COST, ERC, ERDF, Horizon 2020 and Marie Sk{\l}odowska-Curie Actions, European Union;
Investissements d'Avenir Labex, Investissements d'Avenir Idex and ANR, France;
DFG and AvH Foundation, Germany;
Herakleitos, Thales and Aristeia programmes co-financed by EU-ESF and the Greek NSRF, Greece;
BSF-NSF and MINERVA, Israel;
Norwegian Financial Mechanism 2014-2021, Norway;
NCN and NAWA, Poland;
La Caixa Banking Foundation, CERCA Programme Generalitat de Catalunya and PROMETEO and GenT Programmes Generalitat Valenciana, Spain;
G\"{o}ran Gustafssons Stiftelse, Sweden;
The Royal Society and Leverhulme Trust, United Kingdom.
 
The crucial computing support from all WLCG partners is acknowledged gratefully, in particular from CERN, the ATLAS Tier-1 facilities at TRIUMF (Canada), NDGF (Denmark, Norway, Sweden), CC-IN2P3 (France), KIT/GridKA (Germany), INFN-CNAF (Italy), NL-T1 (Netherlands), PIC (Spain), ASGC (Taiwan), RAL (UK) and BNL (USA), the Tier-2 facilities worldwide and large non-WLCG resource providers. Major contributors of computing resources are listed in Ref.~\cite{ATL-SOFT-PUB-2021-003}.


\clearpage
 
\printbibliography
 
\clearpage
 
\begin{flushleft}
\hypersetup{urlcolor=black}
{\Large The ATLAS Collaboration}

\bigskip

\AtlasOrcid[0000-0002-6665-4934]{G.~Aad}$^\textrm{\scriptsize 102}$,
\AtlasOrcid[0000-0002-5888-2734]{B.~Abbott}$^\textrm{\scriptsize 120}$,
\AtlasOrcid[0000-0002-1002-1652]{K.~Abeling}$^\textrm{\scriptsize 55}$,
\AtlasOrcid[0000-0001-5763-2760]{N.J.~Abicht}$^\textrm{\scriptsize 49}$,
\AtlasOrcid[0000-0002-8496-9294]{S.H.~Abidi}$^\textrm{\scriptsize 29}$,
\AtlasOrcid[0000-0002-9987-2292]{A.~Aboulhorma}$^\textrm{\scriptsize 35e}$,
\AtlasOrcid[0000-0001-5329-6640]{H.~Abramowicz}$^\textrm{\scriptsize 151}$,
\AtlasOrcid[0000-0002-1599-2896]{H.~Abreu}$^\textrm{\scriptsize 150}$,
\AtlasOrcid[0000-0003-0403-3697]{Y.~Abulaiti}$^\textrm{\scriptsize 117}$,
\AtlasOrcid[0000-0002-8588-9157]{B.S.~Acharya}$^\textrm{\scriptsize 69a,69b,m}$
% ...
\bigskip
\\

% ...
$^{29}$Physics Department, Brookhaven National Laboratory, Upton NY; United States of America.\\
% ...
$^{35}$$^{(a)}$Facult\'e des Sciences Ain Chock, R\'eseau Universitaire de Physique des Hautes Energies - Universit\'e Hassan II, Casablanca;$^{(b)}$Facult\'{e} des Sciences, Universit\'{e} Ibn-Tofail, K\'{e}nitra;$^{(c)}$Facult\'e des Sciences Semlalia, Universit\'e Cadi Ayyad, LPHEA-Marrakech;$^{(d)}$LPMR, Facult\'e des Sciences, Universit\'e Mohamed Premier, Oujda;$^{(e)}$Facult\'e des sciences, Universit\'e Mohammed V, Rabat;$^{(f)}$Institute of Applied Physics, Mohammed VI Polytechnic University, Ben Guerir; Morocco.\\
% ...
$^{49}$Fakult\"{a}t Physik , Technische Universit{\"a}t Dortmund, Dortmund; Germany.\\
% ...
$^{55}$II. Physikalisches Institut, Georg-August-Universit\"{a}t G\"ottingen, G\"ottingen; Germany.\\
% ...
$^{69}$$^{(a)}$INFN Gruppo Collegato di Udine, Sezione di Trieste, Udine;$^{(b)}$ICTP, Trieste;$^{(c)}$Dipartimento Politecnico di Ingegneria e Architettura, Universit\`a di Udine, Udine; Italy.\\
% ...
$^{102}$CPPM, Aix-Marseille Universit\'e, CNRS/IN2P3, Marseille; France.\\
% ...
$^{117}$Department of Physics, New York University, New York NY; United States of America.\\
% ...
$^{120}$Homer L. Dodge Department of Physics and Astronomy, University of Oklahoma, Norman OK; United States of America.\\
% ...
$^{150}$Department of Physics, Technion, Israel Institute of Technology, Haifa; Israel.\\
$^{151}$Raymond and Beverly Sackler School of Physics and Astronomy, Tel Aviv University, Tel Aviv; Israel.\\
% ...
$^{a}$ Also Affiliated with an institute covered by a cooperation agreement with CERN.\\
$^{b}$ Also at An-Najah National University, Nablus; Palestine.\\
$^{c}$ Also at Borough of Manhattan Community College, City University of New York, New York NY; United States of America.\\
$^{d}$ Also at Center for High Energy Physics, Peking University; China.\\
$^{e}$ Also at Center for Interdisciplinary Research and Innovation (CIRI-AUTH), Thessaloniki; Greece.\\
$^{f}$ Also at Centro Studi e Ricerche Enrico Fermi; Italy.\\
% ...
$^{m}$ Also at Department of Physics, King's College London, London; United Kingdom.\\
% ...

\end{flushleft}


\end{document}